\documentclass[amssymb, aps, amsmath, prb, twocolumn]{revtex4-2}
\usepackage{dcolumn}
\usepackage{amsmath}
\usepackage{amssymb}
\usepackage{physics}
\usepackage[utf8]{inputenc}
\usepackage{grffile}
\usepackage{hyperref}
\usepackage{bm}
\usepackage{bbm}
\usepackage{bbold}
\usepackage{listings}
\usepackage{caption}
\usepackage{placeins}
\usepackage{tcolorbox}
\usepackage{booktabs}
\usepackage{tabularx}
\usepackage{pbox}
\usepackage{multirow}
\usepackage{subcaption}
\usepackage{chemformula}
\usepackage{siunitx}
\usepackage{upgreek}
\usepackage{appendix}

\begin{document}
% Title Page
\title{A method to derive material-specific model descriptions of spin degrees of freedom coupled to
    a fermionic bath for materials displaying prevalent spin physics}
\author{Benedikt M. Schoenauer}
\email{benedikt.schoenauer@quantumsimulations.de}
\affiliation{HQS Quantum Simulations GmbH, Rintheimer Straße 23, 76131 Karlsruhe, Germany}
\author{Nicklas Enenkel}
\affiliation{HQS Quantum Simulations GmbH, Rintheimer Straße 23, 76131 Karlsruhe, Germany}
\author{Florian G. Eich}
\affiliation{HQS Quantum Simulations GmbH, Rintheimer Straße 23, 76131 Karlsruhe, Germany}
\author{Vladimir V. Rybkin}
\affiliation{HQS Quantum Simulations GmbH, Rintheimer Straße 23, 76131 Karlsruhe, Germany}
\author{Michael Marthaler}
\affiliation{HQS Quantum Simulations GmbH, Rintheimer Straße 23, 76131 Karlsruhe, Germany}
\author{Sebastian Zanker}
\affiliation{HQS Quantum Simulations GmbH, Rintheimer Straße 23, 76131 Karlsruhe, Germany}
\author{Peter Schmitteckert}
\email{peter.schmitteckert@quantumsimulations.de}
\affiliation{HQS Quantum Simulations GmbH, Rintheimer Straße 23, 76131 Karlsruhe, Germany}

\date{\today}

\begin{abstract}
Magnetism and spin physics are true quantum mechanical effects and their description usually requires
multi reference methods and is often hidden in the standard description of molecules in quantum chemistry.
In this work we present a twofold approach to the description of spin physics in molecules and
solids.
First, we present a method that identifies the single-particle basis in which a given subset of the orbitals
is equivalent to spin degrees of freedom for models and materials which feature significant spin
physics at low energies.
We introduce a metric for the spin-like character of a basis orbital, of which the
optimization yields the basis containing the optimum spin-like basis orbitals.
Second, we demonstrate an extended Schrieffer-Wolff transformation method to derive the effective Hamiltonian
acting on the subspace of the Hilbert space in which the charge degree of freedom of electron
densities in the spin-like orbitals is integrated out.
The method then yields an effective Hamiltonian describing spins coupled to a fermionic
environment.
This extended Schrieffer-Wolff transformation is applicable to a wide range of Hamiltonians and
has been utilized in this work for model Hamiltonians as well as the Hamiltonian describing the
active orbital space of molecular chromium bromide. This is achieved by reformulating the highly non-linear
Schrieffer-Wolff equations into a linear set of equations corresponding to an operator basis.
\end{abstract}
\maketitle

\section{Introduction}
The theoretical study of the physics of many-body quantum systems has remained notoriously difficult
on conventional computers not least due to the intractable exponential growth of the corresponding Hilbert
spaces. This holds true already for simple model Hamiltonians, but becomes even more problematic for the study
of real materials due to the vast number of degrees of freedom and interactions. To investigate
certain aspects of these materials it is therefore imperative to either utilize significant approximations or
to obtain a much more concise Hamiltonian focused on accurately describing this aspect. In this
article, we concentrate on the description of the spin physics of actual materials. The spin physics is relevant for the
understanding of the material's magnetic properties~\cite{heisenberg_1928} as well as their experimental identification with spin
resonance spectroscopy techniques~\cite{Gunt2013, Lund2011}.
The quantum simulation of spin system Hamiltonians is also perfectly suited as an
application of quantum computing
in general~\cite{McAr2018} and analogue quantum computing in particular\cite{Pouse2023}.

We present a method for the derivation of the effective Hamiltonian description for the relevant spin degrees of
freedom of an actual material embedded in an environment of fermions. The first part of the method
entails the identification of potential relevant orbital spin degrees of freedom. We propose a
new metric for the classification of orbital spin degrees of freedom and present a systematic way
to determine the orbital basis that contains the best realizations of these spins, see also ~\cite{Shirazi2024}.
Given the presence of
relevant spins, we proceed with the second part of the method, a Schrieffer-Wolff transformation
procedure~\cite{Schr1966, Fold1950} aimed at decoupling the
subspace of the Hilbert space containing these spin degrees of freedom from the remaining high-energy
subspaces containing the charge degrees of freedom. Many different Schrieffer-Wolff and related unitary transformation schemes have been
proposed, comprising perturbative~\cite{Schr1966, Brav2011, Landi2024}, variational~\cite{Wurt2019}, and
continuous~\cite{Wegn1994, Krul2012, Schm2022} variants. So
far, the application of these methods has been limited to concise model Hamiltonians. We present an
extended version of the perturbative single step Schrieffer-Wolff transformation that can be applied to generic
Hamiltonian descriptions of materials, given the presence of relevant low-energy spin physics in the
material. By restricting ourselves to spin-like orbitals that have by definition no or only small charge fluctuations
we ensure that a perturbative treatment of the charge fluctuations is justified.
This is also a main difference to canonical transformation approaches \cite{White:2002} where on tries achieve a canonical diagonalization.
We restrict our canonical transformation to a perturbative regime where the application does not create large higher order terms.

We demonstrate the accuracy of the effective spin-bath model Hamiltonians,
where the spin-like orbitals are treated as spins coupled to a fermionic bath,
resulting from our Schrieffer-Wolff transformation for the description
of the low-energy physics both for simple model Hamiltonians with
established results~\cite{Schr1966, MacD1988} as well as actual
materials. We further discuss the extent to which the quality of the effective model
Hamiltonian for the spins coupled to the fermionic environment derived with the proposed method can
be anticipated.

We would like to point out that we do start with an approximate multi reference calculation to determine
a suitable single particle basis. However, the Schrieffer-Wolff expansion, as explained below, starts from
the full quantum chemical description only using the adapted basis set from the multi reference calculation.
At first glance this may appear as a waste of resources. However, as shown below,
already starting in a suitable single particle basis leads to a reasonably good spin-bath model, which then improved
by the Schrieffer Wolff transformation.
This approach is in strong contrast to a traditional setup, where one tries to extract either an effective Hubbard
model using DFT+U \cite{Anisimov:1991,Marzari2024}, DFT+DMFT \cite{RevModPhys.68.13,Kotliar2006,Vollhardt:2019}
or constrained RPA \cite{arya2006, Springer:1998, RPA:2004,Jepsen:2010,Hohnerkamp:2018}
in the framework of maximally localized Wannier functions \cite{Wannier:1937,Kohn:1959,RevModPhys.84.1419},
which is can then be transformed into a spin-math model.
Or alternatively using DFT based approaches using the magnetic force theorem \cite{Liechtenstein:1987,TB2J:2021}.
In these approaches various approximation are staggered. In our work we only rely on the Schrieffer-Wolff transformation.

\section{Methodology}
In Section~\ref{seq:spin_like} we define the notion of spin-like orbitals and in
Section~\ref{seq:local_parity} we propose the local
parity as a metric for the spin-like character of orbitals. In Section~\ref{seq:opt_parity} we present our parity
optimization procedure as a way to determine spin-like orbitals, as well as a way to
gauge the presence of relevant low energy spin physics in the system. We discuss the
basic concept of the Schrieffer-Wolff transformation in Section~\ref{sec:base_sw}.
We provide  our proposed extended Schrieffer-Wolff transformation method in detail
in Sections~\ref{sec:sym_spec} and~\ref{seq:sw_lgs}. In
Section~\ref{sec:workflow} we show how
the described methods are combined into our workflow for deriving effective model Hamiltonians of
spins in a fermionic environment
for materials with relevant low-energy spin physics.
\subsection{Spin-like orbitals}
\label{seq:spin_like}
We consider an orbital $\phi_i$ as being spin-like only if the electron density $n_i$ contained in it is
strictly equal to one. This requires that the average electron density in the orbitals $\phi_i$
satisfies $\langle n_i \rangle = 1$. It furthermore requires that the fluctuations around this average
electron density satisfy $\delta n_i \rightarrow 0$. Negligible fluctuations around the average
electron density imply that electron density of the orbital does not affect the low-energy
dynamics of the system and vice versa. Just an average electron density $\langle n_i \rangle = 1$
places no restrictions on the orientation and dynamics of the electron spin in the orbital $\phi_i$. When
both requirements are simultaneously met by the states in the low-energy Hilbert space, the
dynamics of the electron density in the orbital $\phi_i$, often referred to as the charge degree of
freedom, becomes superfluous to the description of the dynamics of the system. The electron density
in the orbital $\phi_i$ consequently couples to the remainder of the system exclusively via its spin degree of
freedom. It is then sufficient to represent the degrees of freedom of the electron
density contained in the orbital $\phi_i$ as a pure spin degree of freedom and to employ the
associated spin operator algebra. The local Hilbert space $\mathcal{H}_i$ of a spin degree of freedom is half the size
of the local Hilbert space of a fermionic orbital and is furthermore naturally represented on a
Qubit. We therefore aim to determine each spin-like basis orbital or linear combinations thereof
meeting both stated requirements, so that they can be represented as spins.

\subsection{Local parity}
\label{seq:local_parity}
We propose the ground state local parity $P_i$ as a measure for the spin-like character of an orbital
$\phi_i$. The operator representation of the local parity reads
\begin{align}
    P_i = (-1)^{n_{i\uparrow} + n_{i\downarrow}}\,,
\end{align}
where $n_{i\uparrow} = c^{\dagger}_{i\uparrow} c^{}_{i\uparrow}$ denotes the electron density in
$\phi_i$ with electron spin quantum number $s^z_i = +1/2$ and $n_{i\downarrow}=
c^{\dagger}_{i\downarrow} c^{}_{i\downarrow}$ the electron density with quantum number
$s^z_i = -1/2$ respectively, where we are using units with $\hbar=1$.
The local Hilbert space $\mathcal{H}_i$ of the orbital $\phi_i$ is spanned by the states,
\begin{align}
    \left\lbrace \vert 0 \rangle,\, \vert \uparrow \rangle,\, \vert \downarrow \rangle,\, \vert \uparrow\downarrow \rangle \right\rbrace\,,
\end{align}
and the action of the local parity operator on these states reads
\begin{align}
\begin{array}{lcl}
    P_i \, \vert 0 \rangle & = & +1\, \vert 0 \rangle \\
    P_i \, \vert \downarrow \rangle & = & -1\, \vert \downarrow \rangle \\
    P_i \, \vert \uparrow \rangle & = & -1\, \vert \uparrow \rangle \\
    P_i \, \vert \uparrow\downarrow \rangle & = & +1\, \vert \uparrow\downarrow \rangle \,.
\end{array}
\end{align}
For states where the orbital contains a single electron the local parity operator $P_i$ returns the eigenvalue
$p_i = -1$. For the remaining two basis states $P_i$ returns the eigenvalue $p_i = +1$. Any state $\vert \psi \rangle$ in the local
Hilbert space $\mathcal{H}_i$ that contains contributions from non singly occupied basis states hence
satisfies
\begin{align}
    P_i \vert \psi \rangle = \left(-1 + \alpha\right) \vert \psi \rangle \,,
\end{align}
with $2 \geq \alpha > 0$, because the resulting fluctuations in the electron density $\delta n_i
\neq 0$ manifest
themselves in strictly positive contributions to the local parity.
An alternative and more useful operator representation of the local parity
reads
\begin{align}
    P_i =& \left(1 - 2 c^{\dagger}_{i\uparrow} c_{i\uparrow}\right) \left(1 -2
    c^{\dagger}_{i\downarrow} c_{i\downarrow}\right)\\\nonumber
    =& 1 - 2\left(c^{\dagger}_{i\uparrow}c_{i\uparrow} + c^{\dagger}_{i\downarrow} c_{i\downarrow}\right) + 4\, c^{\dagger}_{i\uparrow}c^{\dagger}_{i\downarrow}c_{i\downarrow}c_{i\uparrow} \,,
\end{align}
and its corresponding expectation value with respect to the many-body state $\vert n \rangle$ can be
expressed as
\begin{align}
    \langle n \vert P_i \vert n \rangle = 1 - 2 \rho^{(1)}_{ii} + 4 \rho^{(2)}_{iiii} \,,
\end{align}
where
\begin{align}
    \rho^{(1)}_{qp} = \sum_{\sigma} \langle n \vert c^{\dagger}_{q\sigma} c^{}_{p\sigma} \vert n \rangle\,,
\end{align}
denotes the one-electron reduced density matrix (1-RDM) and
\begin{align}
    \rho^{(2)}_{qprs} = \langle n \vert c^{\dagger}_{q\uparrow} c^{\dagger}_{p\downarrow}c^{}_{r\downarrow}
    c^{}_{s\uparrow} \vert n \rangle\,,
\end{align}
denotes the two-electron reduced density matrix (2-RDM) respectively. If we choose $\vert n \rangle
= \vert \psi_0 \rangle$, with $\vert \psi_0 \rangle$ a good approximation of the ground state of the
system, we can identify orbitals $\phi_i$ for which $\langle P_i \rangle_0 = \langle \psi_0 \vert P_i
\vert \psi_0 \rangle = -1 + \varepsilon$, with $\varepsilon \rightarrow 0$, as spin-like orbitals of
the system. In general, the spin-like orbitals of the system do not coincide with the basis
orbitals typically used in standard basis sets of quantum chemistry codes.
We therefore require a method to determine the set $\lbrace \phi_i \rbrace$ of orthonormal linear combinations of
basis orbitals, for which the local parities most closely approach $\lbrace p_i \rbrace = -1$.

\subsection{Local parity optimization of the orbital basis}
\label{seq:opt_parity}
We propose an iterative procedure to determine the particular orbital basis in which the local
parities are extremal.
For this we attempt a sequence of unitary pairwise rotations of the orbital fermionic operators given by
\begin{align}
    c_{q\sigma} &= \cos \theta \, c_{i\sigma} + \sin \theta \, c_{j\sigma}\,, \nonumber\\
    c_{p\sigma} &= -\sin \theta \, c_{i\sigma} + \cos \theta \, c_{j\sigma}\,,
\end{align}
with the same rotation being performed for the Hermitian conjugates of the operators.
From the reduced density matrices $\rho^{(1)}$ and $\rho^{(2)}$ we can compute the local parity of
an orbital $\phi_q$, which results from the linear combination of orbitals $\phi_i$ and $\phi_j$, as
$\langle P_q \rangle_0 (\theta)$.
The local parity $\langle P_{q}\rangle_0(\theta)$ is an analytic, $2\pi$-periodic function of the rotation angle
$\theta$. We find the extremal points $\theta_n$ of the function $\langle P_{q} \rangle_0 (\theta)$ in
the domain $\theta \in [0, 2 \pi)$ from
\begin{align}
    \label{eq:cond_rot}
    \left. \frac{d\langle P_{q} \rangle_0}{d\theta}\right\vert_{\theta_n} = 0\,,
\end{align}
and select solutions $\theta_n$ that satisfy
\begin{align}
    \left . \frac{d^2 \langle P_{q} \rangle_0}{d\theta^2}  \right\vert_{\theta_n} \neq 0 \,.
\end{align}
The analytic expression for the derivatives of the function $\langle P_{q} \rangle_0 (\theta)$ are
shown in appendix~\ref{sec:diff_p}.
If a solution $\theta_m$ exists which satisfies $d^2\langle P_{q} \rangle_0 / d\theta^2 \vert_{\theta_m} >
0$ and $\langle P_{q} \rangle_0 (\theta_m) \leq \langle P_{q}\rangle_0 (\theta_l)\, \forall \theta_l \in \theta_n$, we
accept the rotation attempt $i\rightarrow q$ and $j\rightarrow p$ with rotation angle $\theta_m$ and
we reject the rotation attempt otherwise.
By repeating the procedure for each pair of basis orbitals $\phi_i$ and $\phi_j$, we arrive at
at an orthonormal basis in which the local parities have taken up extremal values.
We then identify the orbitals $\phi_q$ of the resulting basis for which $\langle
P_{q} \rangle_0 = -1 + \varepsilon$, with $\varepsilon$ an arbitrary small, positive value, as the
spin-like orbitals of the system. Basis orbitals $\phi_p$ with a local parity $\langle P_p\rangle_0 \simeq +1$ can be
regarded as beneficial for the purpose of separating the system's spin degrees of freedom from their respective
environment since they experience exclusively a transfer of an even number of particles only, such that the spin
degree of freedom of electrons occupying the orbitals $\phi_p$ becomes insignificant.
If the optimization gets stuck we found that first optimizing for minimal parities, and then optimizing the non spin-like orbitals
for maximal parity we get out off local minima.
In addition, we have implemented an optimization scheme using the full gradient.
Currently, the pairwise optimization is typically more efficient, but this approach may help in the future to improve convergence.
\subsection{Perturbative similarity transformation}
\label{sec:base_sw}
In principle, any Hamiltonian can be completely diagonalized by means of a particular unitary transformation $U$.
In practice, finding the particular unitary transformation $U$ often requires a complete diagonalization of the Hamiltonian to begin with.
Here we recap how a perturbative similarity transformation, namely the Schrieffer-Wolff
transformation~\cite{Schr1966, Brav2011}, can be used to determine an approximate transformation operator $U$, or the generator
$S$ thereof, which does not diagonalize the Hamiltonian fully, but yields a block-diagonal
Hamiltonian instead.
These blocks consist of the orthonormal states in the Hilbert space $\mathcal{H}$ that share a
given choice of characteristics, e.g.~the local particle quantum number $n_i$.
We will denote the set of terms in the Hamiltonian that are already block-diagonal in the intial basis as $H_0$.
The remaining terms connect different blocks, i.e~are block-offdiagonal, and are denoted $V$. The
complete Hamiltonian thus reads
\begin{align}
    H = H_0 + V\,.
\end{align}
A unitary similarity transformation of the Hamiltonian is given by
\begin{align}
    \tilde{H} = U^{\dagger} H U = \text{e}^{-S} \left(H_0 + V\right) \text{e}^{S} = \text{e}^{-S}H_0
    \text{e}^{S} + \text{e}^{-S} V \text{e}^{S}\,,
\end{align}
where $S$ is an anti-Hermitian operator. One refers to it as the generator of the Schrieffer-Wolff transformation.
The key problem of the Schrieffer-Wolff transformation becomes finding the generator $S$ such that
the transformed Hamiltonian $\tilde{H}$ becomes entirely block-diagonal.
In order to arrive at an equation for $S$ one makes use of the Campbell-Baker-Hausdorff formula to expand
\begin{widetext}
\begin{align}
    \tilde{H} = \mathrm{e}^{S} H \mathrm{e}^{-S} = \sum_{m=0}^{\infty} \frac{1}{m!}[S,\, H]_m = H + [S,\, H] + \frac{1}{2}\left[S,\, [S,\, H]\right] + \dots\,,
    \label{eq:sw_series}
\end{align}
\end{widetext}
where for generators $S$ satisfying $\Vert S \Vert \ll \Vert H_0 \Vert$, with $\Vert \cdot \Vert$ a suitable norm, one can approximate the expression as
\begin{align}
    \tilde{H}= H_0 + V + \left[S,\, H_0 \right] + \left[S,\, V\right] + \mathcal{O}(S^2)\,.
    \label{eq:sw_approx}
\end{align}
Considering that commutators of pairs of block-diagonal operators or pairs of block-offdiagonal
operators respectively generally become block-diagonal, while the commutators of block-diagonal
operators with block-offdiagonal operators become block-offdiagonal, one chooses the equation
\begin{align}
    [S,\, H_0] = -V\quad \Leftrightarrow \quad [H_0,\, S] = V\,,
    \label{eq:sw_cond}
\end{align}
by which one can determine the generator $S$ which removes the block-offdiagonal terms $V$ of the
Hamiltonian. If a solution $S$ to eq.~(\ref{eq:sw_cond}) exists, one can use
\begin{align}
    \left[S, \left[S,\, H_0\right]\right] = \left[S,\, -V\right] = -\left[S,\, V\right] \not\in
    \mathcal{O}(S^2)\,,
\end{align}
to simplify the expression for the transformed Hamiltonian
\begin{align}
    \tilde{H} = \mathrm{e}^{-S} \left(H_0 + V\right) \mathrm{e}^{S}  \simeq H_0 + \frac{1}{2}\left[S,\, V\right] + \mathcal{O}(S^2)\,,
\end{align}
where the terms originating from $[S,\, V]/2$ contain the
perturbative corrections arising from the consecutive application of two block-offdiagonal operators.
A subsequent projection to the subspaces
\begin{align}
    \mathcal{P}_n = \sum_{p\in \mathcal{P}_n} \vert p \rangle \langle p \vert\,,
    \label{eq:h_proj}
\end{align}
yields the block-diagonal Hamiltonian
\begin{align}
    \tilde{H}_{\mathrm{block-diagonal}} = \sum_{n} \mathcal{P}_n \tilde{H} \mathcal{P}_n\,,
    \label{eq:h_block_diagonal}
\end{align}
where $n$ denotes the distinct blocks of the Hilbert space.

\subsection{Symmetry specification of block-offdiagonal operators}
\label{sec:sym_spec}
The block-offdiagonal part of the Hamiltonian $V$ consists of a sum of block-offdiagonal terms. These in turn
comprise products of individually block-offdiagonal operators $x$. In the following we detail a
method to decompose generic block-offdiagonal operators $x$ into distinct components.
Each of the components exclusively connects two distinct blocks of the Hilbert space $\mathcal{H}$,
often associated with distinct quantum numbers of a symmetry of the system.
A given block-offdiagonal operator $x: \mathcal{H} \rightarrow \mathcal{H}$
satisfies
\begin{align}
    \left[H_0,\, x\right] = \varepsilon z\neq 0\,,
\end{align}
where $\varepsilon$ denotes an arbitrary scalar and $z: \mathcal{H}\rightarrow \mathcal{H}$ an arbitrary operator.
Let $A$ be a diagonal operator in the initial basis. It can be identical
to the symmetry operator differentiating the blocks of the Hilbert space, but is not required to be.
One can use the spectrum of $A$ to expand the operator $x$ as
\begin{align}
    x =\sum_{q} x_q = \sum_{q} \beta_q \prod_{i \neq q} \left(A - a_i\right) x\,,
    \label{eq:operator_splitting}
\end{align}
where the different $x_q$ couple the target subspace associated with the eigenvalue
$a_q$ of $A$ to other subspaces of the Hilbert space. If the operator $A$ satisfies
\begin{align}
    [x, A] = 0\,,
\end{align}
then both subspaces, initial and final, coupled via $x_q$ are specified by the eigenvalue $a_q$. This is
possible for the fermionic creation and annihilation operators $c^{\dagger}_{i\sigma}$ and $c_{i\sigma}$
and is displayed in appendix~\ref{sec:sym_ops}.
The coefficients $\beta_q$ are solutions to the equation
\begin{align}
    \beta_q \prod_{i \neq q} \left(a_q - a_i\right) = 1\,.
    \label{eq:splitting_prefactor}
\end{align}
The symmetry-specified block-offdiagonal operators $x_q$ satisfy
\begin{align}
    \left[x_q,\, x^{\dagger}_{q'}\right] \propto \delta_{qq'}\,,
\end{align}
where the operator $x^{\dagger}_{q}$ denotes the Hermitian conjugate of the operator $x_q$.

\subsection{Schrieffer-Wolff transformation as a system of linear equations for unique
block-offdiagonal operators}
\label{seq:sw_lgs}
Following the procedure outlined in Section~\ref{sec:base_sw} we separate the Hamiltonian of the
system into its block-diagonal and block-offdiagonal contributions as
\begin{align}
    H = H_0 + V\,.
\end{align}
The block-offdiagonal contribution $V$ comprises each block-offdiagonal term $v$
\begin{align}
    V =& \sum_{\left\lbrace v \right\rbrace} \alpha_{v} \left[\left(\prod_{j} o^j \prod_{i}x^i\right)
    + \left(\prod_j o^j \prod_{i} x^i \right)^{\dagger}\right] \\\nonumber
    =& \sum_{\left\lbrace v \right\rbrace} \alpha_{v} \left[\left(\prod_{i,j} o^j \sum_{q(i)}
    x^i_{q(i)}\right) + \left(\prod_{i,j} o^j \sum_{q(i)} x^i_{q(i)} \right)^{\dagger}\right]\,,
\end{align}
where $\prod_i x^i$ denotes sequences of individually block-offdiagonal operators, $\prod_j o^j$
denotes sequences of individually block-diagonal operators, and $x^i_{q(i)}$ denotes the
symmetry-specified components of the operator $x^i$.
We introduce a vector spaces $\mathcal{V}^{h}_0$  and $\mathcal{V}^a_0$, for which each unique pair of Hermitian, or
anti-Hermitian respectively, symmetry-specified
operator sequences in $V$ corresponds to a unique orthonormal basis vector
\begin{align}
    \hat{e}_v =
    \left\lbrace
    \begin{array}{ll}
        \left(\prod_j o^j \prod_{i} x^i_{q(i)}\right) + \left(\prod_j o^j \prod_{i}
        x^i_{q(i)}\right)^{\dagger} &\in \mathcal{V}^{h}_0 \\
        \left(\prod_j o^j \prod_{i} x^i_{q(i)}\right) - \left(\prod_j o^j \prod_{i}
        x^i_{q(i)}\right)^{\dagger} &\in \mathcal{V}^{a}_0
    \end{array}
\right. \,,
\end{align}
where $\mathcal{V}^h$ denotes the Hermitian vector space and $\mathcal{V}^a$ the anti-Hermitian
vector space.
We define the linear map
\begin{align}
    L:
    \begin{array}{l}
    \mathcal{V}^{h}_0 \rightarrow \mathcal{V}^{a}_1\\
    \mathcal{V}^{a}_0 \rightarrow \mathcal{V}^{h}_1
\end{array}\,,
\end{align}
where the action of the linear map on a vector $\hat{e}_v$ is given by
\begin{align}
    L\,\hat{e}_v = \left[H_0,\, \hat{e}_v\right]\,,
\end{align}
where
\begin{align}
    \begin{array}{lcl}
    (L \hat{e}_v) \in \mathcal{V}^{a}_{1} &\mathrm{if} &\hat{e}_v \in V^h_0  \\
    (L \hat{e}_v) \in \mathcal{V}^{h}_{1} &\mathrm{if} &\hat{e}_v \in V^a_0
    \end{array}
    \,,
\end{align}
with
\begin{align}
    \mathrm{dim}\left(\mathcal{V}_1 \right) \geq \mathrm{dim}\left(\mathcal{V}_0\right)\,,
\end{align}
since $\mathcal{V}_1$ includes additional unique operator sequences generated by $[H_0,
\hat{e}_v]$.
In the vector spaces $\mathcal{V}$ we can translate equation~(\ref{eq:sw_cond}) for the generator $S$
of the Schrieffer-Wolff transformation to
\begin{align}
    L\, \vec{S} = \vec{V}\quad \Leftrightarrow\quad \left[H_0,\, S\right] = V\,,
    \label{eq:new_sw_cond}
\end{align}
with $\vec{S} \in \mathcal{V}^a_0$ and $\vec{V} \in \mathcal{V}^h_1$. Determining the generator $S$ of
the Schrieffer-Wolff transformation becomes equivalent to solving the set of linear
equations~(\ref{eq:new_sw_cond}). In general one finds
\begin{align}
    \mathrm{rank}(L) \leq \mathrm{dim}\left(\mathcal{V}_0\right) \leq
    \mathrm{dim}\left(\mathcal{V}_1\right)\,,
\end{align}
and there is consequently no unique solution $\vec{S}$ to the set of equation~(\ref{eq:new_sw_cond}).
It is intuitive that the terms of $V$ approximately bring about transitions between
distinct eigenstates of $H_0$ belonging to different blocks of the Hilbert space. This is reflected
by
\begin{align}
    \left[H_0,\, \hat{e}^h_v\right] = \left(\Delta E_{0,v} \prod_l o^l \right) \hat{e}^a_v \,,
    \label{eq:commutator_map}
\end{align}
where $\Delta E_{0,v}$ denotes the eigenvalue difference between the two eigenstates of $H_0$ and
$\prod_l o^l$ denotes an arbitrary sequence of individually block-diagonal operators.
From equations~(\ref{eq:new_sw_cond}) and~(\ref{eq:commutator_map}) we can approximate
\begin{align}
    \Vert \vec{S} \Vert \approx \sqrt{\sum_{\lbrace v \rbrace}
    \left(\frac{\alpha_v}{\Delta E_{0,v}}\right)^2}\,,
\end{align}
which highlights the necessity for a significant energy gap $\Delta E_{0,v}$ between the separate
subspaces of the Hilbert space coupled by $V$ in
order for the series expansion of $\tilde{H}$ to $\mathcal{O}(S^2)$ to be considered a good approximation.
We consider terms $v$ for which
\begin{align}
\frac{\alpha^2_v}{\vert \Delta E_{0,v} \vert} \geq 1\,,
\end{align}
to be the resonant terms of $V$ which should be retained in the
transformed Hamiltonian $\tilde{H}$. To identify these resonant terms of $V$, we employ a singular value
decomposition (SVD) of the linear map $L$ and arrive at
\begin{align}
    L = U \Sigma W^{\dagger} = U\left(\Sigma_{>} + \Sigma_{<}\right) W^{\dagger} = L_{\mathrm{gapped}} + L_{\mathrm{resonant}}\,,
\end{align}
where $\Sigma_{>}$ comprises the singular values
\begin{align}
    \sigma_i > \frac{1}{N_v} \sum_{\lbrace v \rbrace} \sqrt{\alpha^2_v}\,,
\end{align}
where $N_v$ denotes the number of terms $v$.
We define the Moore-Penrose pseudoinverse of $L$ that acts exclusively on the gapped, i.e.
non-resonant, terms in $V$ as
\begin{align}
    &L^{+}_{\mathrm{gapped}} = \left(W\, \Sigma^{+}_{>}\, U^{\dagger}\right)\,,
\end{align}
where $\Sigma^{+}_{>}$ denotes the pseudoinverse of $\Sigma_{>}$.
In the absence of a unique solution to~(\ref{eq:new_sw_cond}), the closest approximate
solution~\cite{Penr1956} is given by
\begin{align}
    \vec{S} = \left(L^{+}_{\mathrm{gapped}} \vec{V}\right) \in \mathcal{V}^{a}_0\,,
\end{align}
and the resulting transformed Hamiltonian reads
\begin{align}
    \tilde{H} = H_0 + V_{\mathrm{resonant}} + \left[S,\, V_{\mathrm{resonant}}\right] + \frac{1}{2}
    \left[S,\, V_{\mathrm{gapped}}\right]\,,
    \label{eq:approx_sw}
\end{align}
where $V_{\mathrm{gapped}} = [H_0,\, S]$ and $V_{\mathrm{resonant}} = (V - V_{\mathrm{gapped}})$.
The quality of the approximate transformed Hamiltonian $\tilde{H}$ and its respective
$\tilde{H}_{\mathrm{block-diagonal}}$ (see eq.~\ref{eq:h_block_diagonal}) for the description of the low energy
dynamics of a given system is
discussed in Sections~\ref{sec:siam} to~\ref{sec:crbr}.

In summary, we have mapped the non-linear Schrieffer--Wolff equations, as presented in the original work, \cite{Schr1966}
to a linear set of equations, see also Sec.~\ref{sec:siam}.
Without this representation we could not apply the Schrieffer--Wolff approach to
quantum chemistry descriptions including four-index terms.
\subsection{Full workflow}
\label{sec:workflow}
In the following we outline the steps of the workflow that we use to derive an effective spin-bath
model Hamiltonian from a first principles description of a material.
\paragraph{Computation of the required system information}
We start with an ab-initio electronic
structure calculation~\cite{Helg2000} of the material to determine its ground state. The electronic structure method needs to be a
post-Hartree-Fock or a related method, this excludes density functional theory, to capture the effect of correlations in the two-particle
reduced density matrix $\rho^{(2)}$. From the electronic structure calculation we obtain the
single particle basis orbitals and the corresponding one-electron and two-electron integrals  which specify the Hamiltonian description.
For the ground state of the calculation we
compute the one-particle and two-particle reduced density matrices $\rho^{(1)}$ and $\rho^{(2)}$.
\paragraph{Determination of spin-like basis orbitals}
We utilize the reduced density matrices to assign a local parity $\langle P_i \rangle_0$ to the basis
orbitals $\phi_i$. We then perform pairwise rotations of the basis orbitals
to determine the basis in which the local parities of the basis orbitals is
extremized. If there exist optimized basis orbitals $\phi_q$ with $\langle P_q \rangle_0 + 1 <
\varepsilon$, where we typically chose $\varepsilon \leq 10^{-1}$, we proceed with the subsequent steps
of the workflow. If no spin-like orbitals are found, we terminate the workflow, because the absence
of spin-like orbitals also indicates the absence of relevant spin physics in the system under
consideration.
\paragraph{Schrieffer-Wolff transformation of the Hamiltonian}
We use our Schrieffer-Wolff transformation approach to integrate out the valid terms of the Hamiltonian that
modify the electron density in the spin-like orbitals which leads to renormalized couplings of the
electron spins in the spin-like orbitals to the environment. The valid terms are the ones that couple
subspaces of the Hilbert space between which there exists a significant energy gap.
\paragraph{Construction of the effective spin-bath Hamiltonian}
The transformed Hamiltonian $\tilde{H}$ is projected to the particular subspace of the Hilbert space where the
electron density of spin-like orbitals $\phi_q$ is fixed to $n_q\equiv 1$. Utilizing the identity
$n_q=n_{q\uparrow} + n_{q\downarrow}\equiv 1$ fermionic operators acting on the spin-like
operators are substituted with the corresponding spin operators. The resulting Hamiltonian $\mathcal{P} \tilde{H}
\mathcal{P}$ is the effective spin-bath representation of the material.
\paragraph{Representation on a quantum computing device (optional)}
The effective spin-bath model Hamiltonian $\mathcal{P} \tilde{H} \mathcal{P}$ is re-expressed in terms
of the spin operators that are realized on the specific device.

\section{Examples}
Here we present a selection of systems to which we have applied our methods for identifying spin-like
orbitals and for deriving effective spin-bath model Hamiltonians. In Sections~\ref{sec:toy_model}
and~\ref{sec:para_benzyne} we showcase
the spin identification procedure for a lattice model and for a radical molecular system.
In Sections~\ref{sec:siam} and~\ref{sec:dfh} we compare our Schrieffer-Wolff transformation method with
the established results for two well-known model Hamiltonians. In Section~\ref{sec:crbr} we then apply the
complete workflow to molecular chromium bromide and discuss the accuracy of the effective spin-bath
model Hamiltonians.

\subsection{Spin-bath chain model}
\label{sec:toy_model}
\begin{figure}
    \begin{center}
        \includegraphics[width=0.44\textwidth]{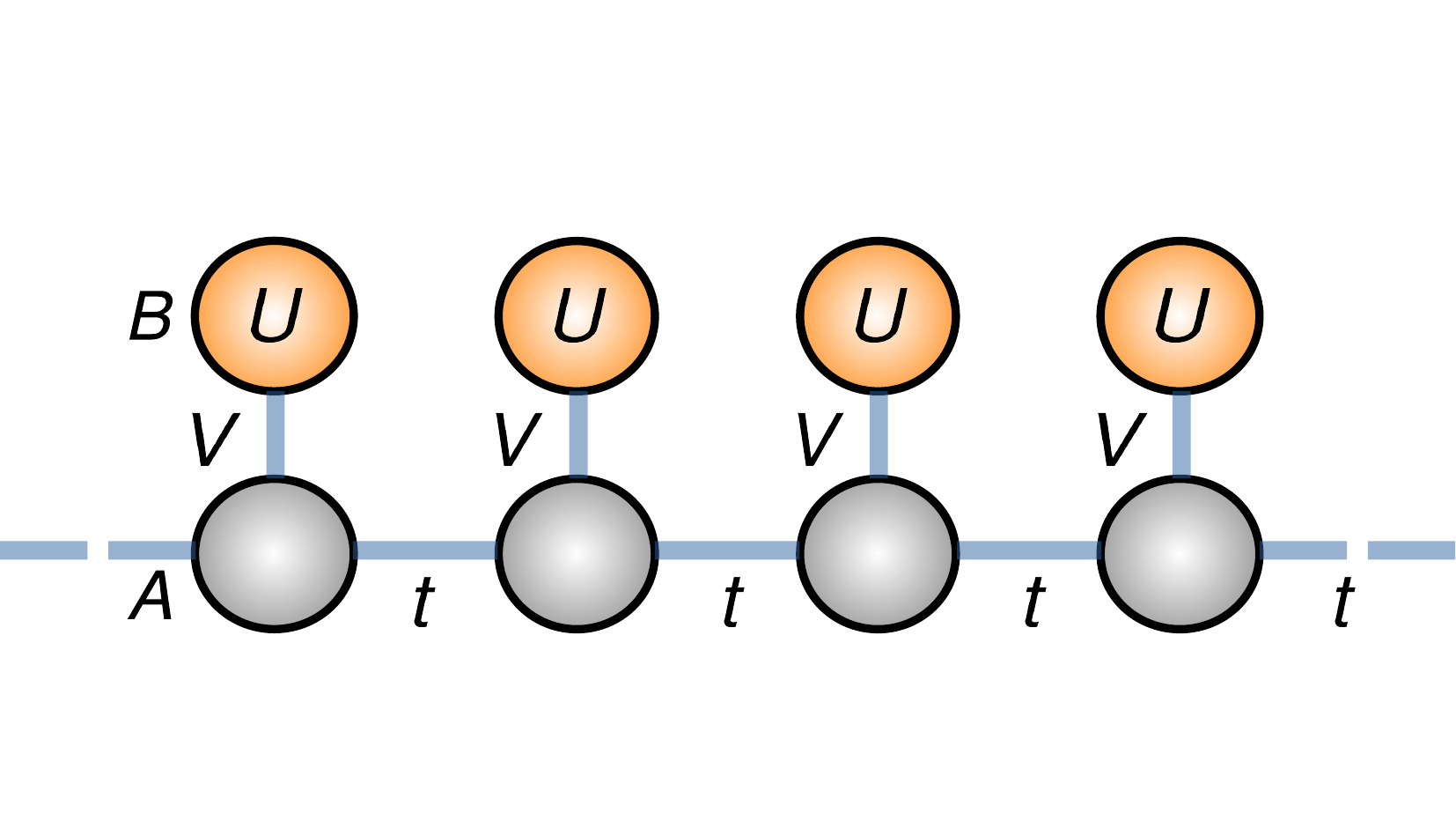}
        \caption{Schematic representation of the lattice model described by the Hamiltonian
            ~(\ref{eq:hamiltonian_toy_model}). The lattice sites $A_i$
    are shown in grey. On the orange lattice sites $B_i$
    the fermions experience a strong repulsive Hubbard interaction $U$ which energetically discourages $n_{B_i}
    \neq 1$. We refer to the lattice sites $A_i$ and $B_i$ as the initial basis of the system.
    The hybridization between the lattice sites $A_i$ and $A_{i+1}$ is given by $t < 0$. We impose periodic boundary
    conditions via a hybridization $t$ between the lattice sites $A_{N=4}$ and $A_1$.
    There is a small hybridization $0 > V > t$ between lattice sites $A_i$ and $B_i$.
    The local Hubbard density-density interaction of strength $U \gg \vert t \vert$ is strongly
    repulsive.
    The connectivity and the coupling constants of the lattice models are chosen such that the
    lattice sites $B_i$ should be good realizations of spin-like orbitals.}
        \label{fig:toy_model_schema}
    \end{center}
\end{figure}
As a first test of the parity optimization procedure we introduce a simple lattice model.
We have chosen the connectivity of the lattice sites and the coupling constants such that for the
lattice sites $B_i$ an electron density $\langle n_{B_i} \rangle_0 \neq 1$ is strongly discouraged.
The lattice sites $B_i$ should consequently be good realizations of spin-like orbitals. The
Hamiltonian of the lattice model reads
\begin{align}
    H =& \sum_{i\sigma} \left(t\, a^{\dagger}_{i+1\sigma} a^{}_{i\sigma} + V\, a^{\dagger}_{i\sigma}
    b^{}_{i\sigma} + \text{h.c.}\right)  \label{eq:hamiltonian_toy_model} \\\nonumber
    &+ U\sum_{i} \left(b^{\dagger}_{i\uparrow} b^{}_{i\uparrow} - \frac{1}{2}\right)
    \left(b^{\dagger}_{i\downarrow} b^{}_{i\downarrow} - \frac{1}{2}\right)\,,
\end{align}
where $a^{\dagger}_{i\sigma}$ creates a fermion of spin $\sigma$ on lattice site $A_i$ and
$b^{\dagger}_{i\sigma}$ creates a fermion of spin $\sigma$ on lattice site $B_i$. The lattice sites
$A_i$ and $B_i$ form the initial basis of the system.
We impose periodic boundary conditions for the lattice sites $A_i$. A graphic representation
of the lattice model is displayed in Fig.~\ref{fig:toy_model_schema}. The coupling constants satisfy
$U \gg \vert t \vert > \vert V \vert$ and $0 > V > t$.
The repulsive Hubbard interaction on lattice sites $B_i$ in combination with the weak hybridization
between $B_i$ and $A_i$ places a significant energy penalty on $n_{B_i} \neq 1$.
Low-energy eigenstates of the Hamiltonian should therefore satisfy $n_{B_i} \equiv 1$ and the
lattice sites $B_i$ be considered good realizations of spin-like orbitals.

We have calculated the low energy spectrum of the Hamiltonian of a chain
of $M=4$ unit cells containing $N=8$ fermions using a numerical diagonalization method.
The reduced density matrices $\rho^{(1)}$ and $\rho^{(2)}$ have been computed for
the corresponding ground state $\vert \psi_0 \rangle$. By definition~\cite{loew1955}, the
eigenstates of $\rho^{(1)}$ form the natural (spin) orbital basis.
We find eigenvalues $\langle n_i \rangle = 1$ of $\rho^{(1)}$ for $N_{\phi_i} = 2$ natural basis
orbitals $\phi_i$. These are considered two natural spin orbitals.
\begin{figure}
    \begin{center}
        \begin{subfigure}{0.49\textwidth}
            \centering
            \includegraphics[width=\textwidth]{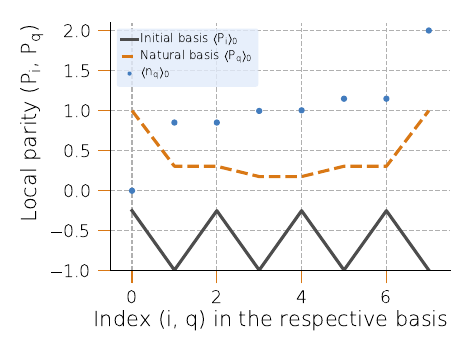}
            \caption{}
            \label{fig:toy_model_no}
        \end{subfigure}
        \begin{subfigure}{0.49\textwidth}
            \centering
        \includegraphics[width=\textwidth]{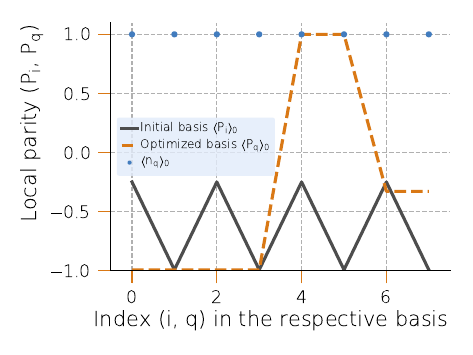}
            \caption{}
            \label{fig:toy_model_po}
        \end{subfigure}
        \caption{Ground state expectation values of the local parity $\langle P_q \rangle_0$ of the initial basis orbitals $A_i$ and
            $B_i$, as well as (a) the natural (spin) orbitals basis and (b) the
            parity optimized basis for the spin-bath chain model described by eq.~(\ref{eq:hamiltonian_toy_model}).
        (a) The dark grey line shows the local parities of the original basis orbitals $A_i$ and
        $B_i$ and the
        dashed orange line indicates the local parities of the natural basis orbitals. The blue
        circles display
        the expectation values for the electron density $\langle n_q \rangle_0$ in the natural
        orbital basis.
        In the natural orbital basis we find $N_{\phi_q}=2$ orbitals $\phi_q$ for which $\langle n_q
        \rangle_0 = 1$,
        but the respective values of the local parities $\langle P_q \rangle_0 > 0$ highlight that $\vert \delta
        n_q\vert \gg 0$.
        (b) The
        dashed orange line now displays the ground state local parities of the parity optimized basis orbitals. The blue
        circles show
        the expectation value for electron density $\langle n_q \rangle_0$ in these optimized basis orbitals.
        In the optimized basis we find $N_{\phi_q} = 4$ orbitals $\phi_q$ with local parity
        $\langle P_q \rangle_0 \simeq -1$. We also observe $\langle n_q \rangle_0 = 1$ for each
        orbital.
        The orbitals $\phi_{q\leq 3}$ of the optimized basis are considered spin-like. They coincide
    with the lattice sites $B_i$, but the ordering has been shuffled in the optimization procedure.}
    \end{center}
\end{figure}
The local parities of the natural orbitals are shown in
Fig.~\ref{fig:toy_model_no}.

We find $\langle P_i \rangle_0 > 0$ for each
natural basis orbital $\phi_i$.
This indicates that the ground state $\vert \psi_0 \rangle$ features significant contributions from
Slater determinants $\vert \xi \rangle$ where $n_i \vert \xi \rangle = 0 \vert \xi \rangle$
or $n_i \vert \xi \rangle = 2 \vert \xi \rangle$ respectively for each natural basis orbital.
The average of these contributions yields $\langle n_i \rangle_0 = 1$,
but the contributions $\langle \psi_0 \vert \xi \rangle \neq 0$ manifest themselves in
strong fluctuations $\vert \delta n_i \vert \gg 0$.
The lattice model highlights that the natural orbitals are generally not suitable candidates for
spin-like orbitals, because they do not satisfy the necessary criterion $\delta n_i \rightarrow 0$.
We point out that the original basis had already contained the spin-like lattice sites $B_i$.
In the given example the natural orbital basis fails to be a good starting point in the
search for spin-like orbitals.
To test the capability of the parity optimization procedure we
have used the natural orbital basis as the starting point, because we have previously found them to be
sufficiently far from the optimal original basis.
The results of the parity optimization are shown in Fig.~\ref{fig:toy_model_po}.

We find $N_{\phi_q}\geq 4$ orbitals $\phi_q$ with local parity $\langle P_q \rangle_0 \simeq -1$.
These optimized orbitals $\phi_q$ coincide with the original lattice sites of $B_i$.
Despite the intentionally difficult starting conditions the parity optimization procedure was able
to recover the original basis featuring the lattice sites $B_i$ that had been designed to
be spin-like.

\subsection{Closed configuration of para-benzyne}
\label{sec:para_benzyne}
\begin{figure}
    \begin{center}
        \begin{subfigure}{0.38\textwidth}
            \centering
            \includegraphics[width=\textwidth]{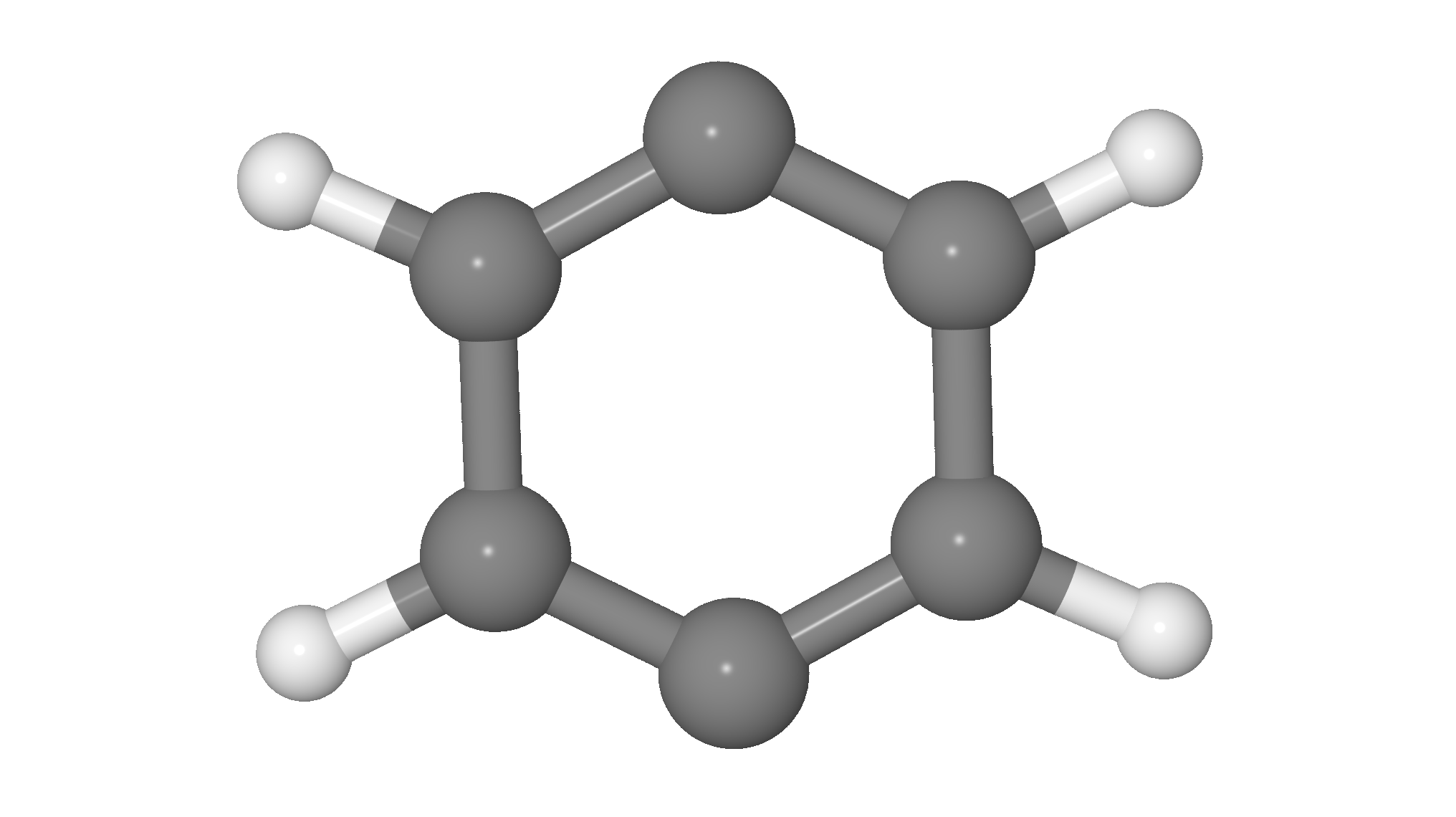}
            \caption{}
            \label{fig:endiine}
        \end{subfigure}
        \begin{subfigure}{0.26\textwidth}
            \centering
            \includegraphics[width=\textwidth]{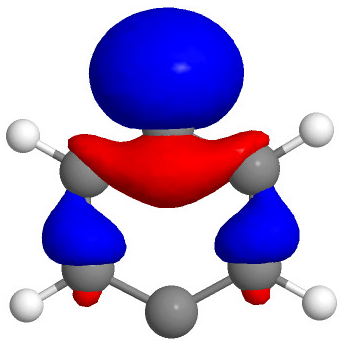}
            \caption{}
            \label{fig:spin_like_1}
        \end{subfigure}
        \begin{subfigure}{0.26\textwidth}
            \centering
            \includegraphics[width=\textwidth]{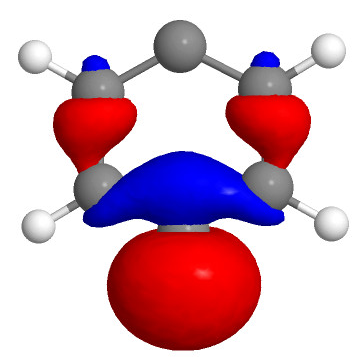}
            \caption{}
            \label{fig:spin_like_2}
        \end{subfigure}
        \caption{(a) Molecular structure of the closed configuration of the molecule para-benzyne
            ($\text{C}_6 \text{H}_4$). (b) Isosurface of one spin-like orbital $\phi_{q=1}$ of
    para-benzyne. (c) Isosurface of the other spin-like orbital $\phi_{q=2}$ of para-benzyne. The
    colors red and blue indicate the sign of the orbital wavefunction.}
    \end{center}
\end{figure}
As a second testbed for the parity optimization procedure we have considered the
molecule para-benzyne $\text{C}_6 \text{H}_4$~\cite{Lind1994}. The molecule is depicted in Fig.~\ref{fig:endiine}.
 In its singlet configuration para-benzyne is known to feature two spin-like orbitals that are
non-trivial linear combinations of the original basis orbitals~\cite{Sale1972}.
The complete active space self-consistent field (CASSCF)~\cite{McWe1957,Roos1980,Helg2000,Lawl2009} method was used to compute the reduced density matrices $\rho^{(1)}$ and $\rho^{(2)}$
in the ground state. CASSCF wave function is constructed as a linear combination of all Slater determinants obtained by distribution of N electrons into M orbitals, which constitute the (N, M) active space, that is $N$ particles on $M$ spinful orbitals are treated within their full Hilbert space, while the remaining orbitals
are treated using a self consistent field approximation  .
Energy is obtained variationally, being minimized wrt. both active orbitals and Slater determinant coefficients.
The orbital basis set comprises $N_{\phi_i}=62$ CASSCF canonical molecular
orbitals of which $N_{\phi_{i,a}}=12$ orbitals are selected as active containing 12 electrons, giving rise to (12, 12) active space.
The Hamiltonian description of the molecule in the molecular orbital basis reads
\begin{align}
    H = \sum_{ij\sigma\sigma'} t^{\sigma \sigma'}_{ij} c^{\dagger}_{i\sigma} c_{j\sigma'} +
    \sum_{ijkl\sigma\sigma'} V^{\sigma\sigma'}_{ijkl} c^{\dagger}_{i\sigma} c^{\dagger}_{j\sigma'}
    c_{k\sigma'} c_{l\sigma}\,,
\end{align}
where the one-electron integrals $t^{\sigma\sigma'}_{ij}$ and two-electron integrals
$V^{\sigma\sigma'}_{ijkl}$ have been determined as part of an ab-initio calculation.
The parity optimization procedure was performed for two sets of initial basis orbitals.
The first set contained all $N_{\phi_i}=62$ basis orbitals of the CASSCF calculation.
The second set was restricted to the active $N_{\phi_i}=12$  orbitals $\phi_i$ with local parity $\langle P_i
\rangle_0 < 9.5\times 10^{-1}$, which coincides with the active space of the CASSCF calculation.
\begin{figure}
    \begin{center}
        %\begin{subfigure}{0.49\textwidth}
        \includegraphics[width=0.49\textwidth]{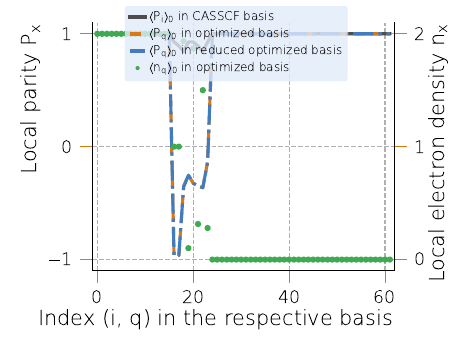}
        \label{fig:endiine_oo}
        \caption{Ground state local parities $\langle P_i\rangle_0$ of the basis orbitals of para-benzyne ($\text{C}_6
            \text{H}_4$) in the initial CASSCF canonical molecular orbitals basis (dark grey line)
            and the
            two different parity optimized bases (dashed orange and blue). The local parities
            $\langle P_i \rangle_0$ of the original atomic basis orbitals each
        satisfy $\langle P_i \rangle_0 > 8.5\times 10^{-1}$, so the initial basis orbitals are not considered spin-like.
    The local parities $\langle P_q \rangle_0$ of the basis orbitals $\phi_q$ in the optimized basis, where the full set of
    $N=62$ orbitals has been optimized, are shown in orange. The resulting local parities $\langle P_q
    \rangle_0$, where the subset of
    $N=12$ basis orbitals $\phi_i$ with initial local parity $\langle P_i \rangle_0 < 9.5\times
    10^{-1}$ has been optimized,
    are displayed in blue. The electron densities in the optimized basis orbitals is shown as green
    circles.
    The restricted set
    turns out to be equivalent to the active space of the CASSCF calculation used to obtain the
    reduced density matrices $\rho^{(1)}$ and $\rho^{(2)}$.
    The local parity $\langle P_q \rangle_0 \simeq -9.4\times 10^{-1}$ of two specific optimized basis orbitals $\phi_q$
    is sufficiently small for the criteria $\langle n_q \rangle_0 \equiv 1$ and $\vert \delta n_q \vert
    \simeq 0$ to be simultaneously fulfilled. The isosurfaces of these spin-like orbitals are
    displayed
    in Figs.~\ref{fig:spin_like_1} and~\ref{fig:spin_like_2}.}
        \label{fig:endiine_parity}
    \end{center}
\end{figure}
The local parities in the initial and the optimized basis are displayed in Fig.~\ref{fig:endiine_parity}.

The differences between the complete and reduced basis sets are insignificant when only considering
the orbitals of smallest local parity after optimization. Discrepancies between the two optimizations
can be observed for basis orbitals of larger local parity. In this case the access to the complete set of
basis orbitals allows for further reduction of the local parity of some basis orbitals in the
optimization.
After parity optimization we identify two basis orbitals $\phi_q$ for which the local parity takes
the value $\langle P_q \rangle_0 \simeq -0.94$. This is sufficiently small a value of the local
parity to be considered spin-like. The local parities of the remaining basis orbitals are
significantly, larger i.e. $\langle P_p \rangle_0 > -0.5$. The two spin-like orbitals $\phi_q$ are
the linear combinations
\begin{align}
    \phi_{q}(\vec{r}) \simeq \frac{1}{\sqrt{2}}\left[\phi_{a}(\vec{r}) \pm \phi_{b}(\vec{r})\right]\,,
\end{align}
of two specific CASSCF canonical molecular orbitals $\phi_a$ and $\phi_b$ of the original basis.
This is consistent with results from previous studies of the closed configuration of
para-benzyne~\cite{Sale1972}.
An isosurface image of the two spin-like orbitals $\phi_q$ is displayed in figures~\ref{fig:spin_like_1} and~\ref{fig:spin_like_2}.

\subsection{Single impurity Anderson model}
\label{sec:siam}
The intuitive first application example for our proposed Schrieffer-Wolff transformation approach is the
single impurity Anderson model (SIAM)~\cite{Ande1961}. It is the model Hamiltonian for which this type of perturbative
similarity transformation was originally
proposed by Schrieffer and Wolff. The generator of the transformation and the transformed
Hamiltonian are both known analytically. The SIAM describes a localized magnetic moment in a system
of non-interacting electrons. The SIAM Hamiltonian reads
\begin{align}
    H =& \sum_{\sigma} \left[-t \sum_{i=1}^{N} \left(c^{\dagger}_{i,\sigma} c_{i-1,\sigma} + \mathrm{h.c.}\right)
    -V_0 \left(d^{\dagger}_{\sigma} c_{0,\sigma} + c^{\dagger}_{0,\sigma} d_{\sigma}\right)\right]
    \\\nonumber
    &+ U \left(d^{\dagger}_{\uparrow} d_{\uparrow} - \frac{1}{2}\right)
    \left(d^{\dagger}_{\downarrow} d_{\downarrow} - \frac{1}{2}\right)\,,
    \label{eq:h_siam}
\end{align}
where $d^{\dagger}_{\sigma}$ creates an electron with spin $\sigma$ on the impurity site and
$c^{\dagger}_{i\sigma}$ creates an electron of spin $\sigma$ on the lattice site $i$ of the remainder of the system.
If one identifies
\begin{align}
    H_0 =& -t\sum_{i=1,\sigma}^{N} \left(c^{\dagger}_{i,\sigma} c_{i-1,\sigma} +
    \mathrm{h.c.}\right)
    \nonumber\\ &+ U \left(d^{\dagger}_{\uparrow} d_{\uparrow} - \frac{1}{2}\right)
    \left(d^{\dagger}_{\downarrow} d_{\downarrow} - \frac{1}{2}\right) \\
    V =& -V_0 \sum_{\sigma} \left(d^{\dagger}_{\sigma} c_{0,\sigma} + c^{\dagger}_{0,\sigma} d_{\sigma}\right)\,,
\end{align}
the canonical transformation by Schrieffer and Wolff~\cite{Schr1966, Kehr1996, HaqS2020} yields the Hamiltonian
\begin{align}
    \tilde{H} =& -\sum_{k,k'\sigma\sigma'} J_{kk'} \left(d^{\dagger}_{\sigma} \vec{\sigma}
    d_{\sigma'}\right) \cdot \left(c^{\dagger}_{k\sigma'}
    \vec{\sigma} c_{k'\sigma}\right)\\\nonumber
    &+ \sum_{k\sigma}\left(W_{kk} + \frac{J_{kk}}{2}
    d^{\dagger}_{\bar{\sigma}} d_{\bar{\sigma}}\right) c^{\dagger}_{k\sigma} c_{k\sigma}\\\nonumber
    &+ \sum_{kk'\sigma} \frac{J_{kk'}}{4} c^{\dagger}_{k\bar{\sigma}} c^{\dagger}_{k'\sigma}
    d_{\sigma} d_{\bar{\sigma}} + \mathrm{h.c.} + \sum_{k\sigma} \varepsilon_k c^{\dagger}_{k\sigma}
    c_{k\sigma}\,,
    \label{eq:h_siam_ana}
\end{align}
where the parameters
\begin{align}
    V_k &\propto V_0\,, \\
    \varepsilon_k &= -2 t \cos k\,,
\end{align}
are the consequence of a Fourier transformation to momentum space
$k=\frac{1\pi}{N+1},\dots,\frac{N \pi}{N+1}$,
while the effective coupling constants
\begin{align}
    J_{k'k} &= V_{k'}V_{k} \left(\frac{1}{\varepsilon_k - \frac{U}{2}} + \frac{1}{\varepsilon_{k'} -
        \frac{U}{2}} - \frac{1}{\varepsilon_k + \frac{U}{2}} - \frac{1}{\varepsilon_{k'} +
    \frac{U}{2}}\right)\,, \\
    W_{k'k} &= \frac{V_{k'} V_k}{2} \left(\frac{1}{\varepsilon_k + \frac{U}{2}} +
    \frac{1}{\varepsilon_{k'} + \frac{U}{2}}\right)\,,
\end{align}
are the result of the Schrieffer-Wolff transformation. A subsequent projection $\mathcal{P}$ of the
Hamiltonian $\tilde{H}$ to
the subspace of the Hilbert space in which the impurity site is singly occupied yields the Kondo
Hamiltonian~\cite{Hews1997}
\begin{align}
    \mathcal{P} \tilde{H} \mathcal{P} = \sum_{k\sigma} \varepsilon_k c^{\dagger}_{k\sigma}
    c_{k\sigma} - \sum_{kk'\sigma\sigma'} J_{k'k}\vec{S}_d \cdot c^{\dagger}_{k'\sigma} \vec{\sigma}
    c_{k\sigma'}\,,
    \label{eq:h_siam_ana_proj}
\end{align}
where the magnetic impurity is coupled to the electrons in the lead via its spin degree of freedom
$\vec{S}_d = (S^x_d, S^y_d, S^z_d)$ only.

\begin{figure}
    \begin{center}
        \includegraphics[width=0.49\textwidth]{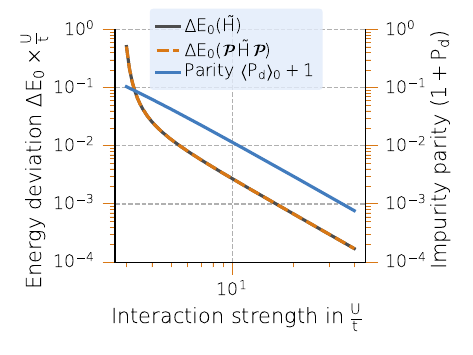}
        \caption{Rescaled ground state energy difference $U\times\Delta E_0/t=U\times\vert E_{0}^{
        (\mathcal{P})\tilde{H}(\mathcal{P})} - E_{0}^{H}\vert /t$ between the ground state of the SIAM Hamiltonian
            and the
        transformed Hamiltonians $\tilde{H}$ and $\mathcal{P} \tilde{H} \mathcal{P}$ as a function
        of the interaction strength $U/t$.
        The dark grey line represents the scaled energy difference for the transformed
        Hamiltonian $\tilde{H}$.
        The dashed orange displays $U\times\Delta E_0/t$
        for the effective Hamiltonian $\mathcal{P} \tilde{H} \mathcal{P}$ where the
        impurity site is restricted to single occupancy $n_d  \equiv 1$.
        The solid blue line displays the local parity $\langle P_d \rangle_0$
        of the impurity site in the true ground state for each value of the interaction strength $U/t$.}
        \label{fig:siam_energy}
    \end{center}
\end{figure}

We have studied a SIAM Hamiltonian of with 9 lattice sites for the lead and an additional site
representing the magnetic impurity and a hybridization between lead and impurity $V_0/t = 1/4$. The
hybridization $t$ between lead lattice sites represents the energy scale of the system. We have
fixed the electron number $N_{\mathrm{electrons}}=10$ and $\sum_i \sigma^z_i=0$ to perform a ground state
computation using a numerical diagonalization method for a range of impurity interaction
strength values $3 \leq U/t \leq 40$. We have calculated the local parity $\langle P_d \rangle_0$ of the
impurity site in the ground state. We have used our Schrieffer-Wolff transformation approach to
numerically determine the Hamiltonians $\tilde{H}$ and $\mathcal{P} \tilde{H} \mathcal{P}$ for each
value of the interaction strength $U/t$ and have compared them with the established and analytic
results~(\ref{eq:h_siam_ana}) and~(\ref{eq:h_siam_ana_proj}). The numerical results match the
analytic expressions exactly. We have also calculated the discrepancy $\Delta E_0 = \vert E_{0}^{
(\mathcal{P})\tilde{H}(\mathcal{P})} - E_{0}^{H}\vert$ between the ground state
energies $E_0$ of the SIAM Hamiltonian $H$~(\ref{eq:h_siam}) and the energy of respective ground states of the
transformed Hamiltonians $\tilde{H}$ and $\mathcal{P}\tilde{H}\mathcal{P}$. The ground state local
parity $\langle P_d \rangle_0$ of the magnetic impurity and the rescaled ground energy discrepancies
$U\Delta E_0 /t$ as a function of the interaction strength $U/t$ are shown in
Fig.~\ref{fig:siam_energy}.

For
values of the interaction strength $U/t \leq 4$, we observe several pronounced energy differences
between the SIAM ground states and the ground states of the transformed and effective Hamiltonians.
In this weak interaction regime, the chemical potential of the impurity lies within the bandwidth $D
= 4t$ of the tight-binding chain representing the remainder of the system. There is then no
energy gap between the subspace of the Hilbert space where the impurity is singly occupied and the
rest of the Hilbert space. In the absence of an energy gap, limiting the series
expansion~(\ref{eq:sw_series}) to first order in $S$ is not a justified approximation. For values of
the interaction strength $U/t > 4$ we observe a decrease of the energy discrepancy $\Delta E_0$
of the ground state energies of the SIAM Hamiltonian and the transformed and effective Hamiltonians
respectively as a power-law $(U/t)^{-\alpha}$ of the interaction strength with an exponent $\alpha > 2$.
The energy discrepancy quickly becomes small with respect to the energy scale
$t$ and we find $\Delta E_0/t \rightarrow 0$ for $U/t \rightarrow \infty$.
For the ground state local parity on the impurity site we observe $7\times 10^{-2} > (\langle P_d \rangle_0 +
1) > 3\times 10^{-2}$ in the regime $U/t \leq 4$. For larger values of the interaction strength $U/t > 4$ we
find that the local parity of the impurity site approximates a polynomial function $(\langle P_d
\rangle_0 + 1)(U/t) \propto (U/t)^{-\beta}$ with the exponent $\beta \approx \alpha -1$. The local
parity of the impurity site appears to be a good indicator of quality of the
approximation~(\ref{eq:sw_approx}) if $\tilde{H}$ is supposed to become block-diagonal in the
subspaces of the Hilbert space with different particle number $n_d$ on the impurity site.
This supports our claim that a local
parity $(\langle P_d \rangle_0 + 1) \rightarrow 0$ signals that the corresponding site can be
represented as a spin degree of freedom and that $\mathcal{P} \tilde{H} \mathcal{P}$ is then a good
effective Hamiltonian to describe the low-energy spin dynamics of the system. The energy discrepancy
$\Delta E_0$ is identical for both $\tilde{H}$ and $\mathcal{P} \tilde{H} \mathcal{P}$. In the case
of the SIAM Hamiltonian the Schrieffer-Wolff transformation yields a Hamiltonian $\tilde{H}$
that does no longer couple the distinct subspaces characterized by $n_d = 1$ and $n_d \neq 1$.
We find that our Schrieffer-Wolff transformation method recovers the analytically known results and
that the ground state of effective Hamiltonian $\mathcal{P} \tilde{H} \mathcal{P}$ approaches the
true ground state as $U/t \rightarrow \infty$ at a rate faster than $(U/t)^{-1}$.

\subsection{Disordered Fermi-Hubbard model}
\label{sec:dfh}
Another significant model Hamiltonian, for which the analytic expression for the Schrieffer-Wolff transformed
Hamiltonian is known, is the Fermi-Hubbard model~\cite{Hubb1963, Essl2005}. It is translation invariant and describes the electrons in orbitals with small
nearest-neighbor hybridizations $t$ where electrons occupying the same orbital exert a density-density
interaction $U$ on each other. In the limit $U/t \rightarrow \infty$ and
$N_{\mathrm{electrons}}/N_{\mathrm{orbitals}} =1$ (half filling) the transformed Hamiltonian $\tilde{H}$ reduces to the
well-known Heisenberg model of interacting spins~\cite{MacD1988, Wurt2019}. Here, we present the Schrieffer-Wolff
transformation for the disordered Fermi-Hubbard (dFH) chain. Its Hamiltonian is given by
\begin{align}
    \label{eq:h_dis_hubbard}
    H =& \sum_{i=1}^N \sum_{\sigma}\left[\varepsilon_i c^{\dagger}_{i,\sigma} c_{i,\sigma}
    -t\left(c^{\dagger}_{i,\sigma} c_{i-1,\sigma} + \mathrm{h.c.}\right)\right] \\\nonumber
    &+ U \sum_{i=1}^N \left(c^{\dagger}_{i\uparrow} c_{i\uparrow} - \frac{1}{2}\right)
    \left(c^{\dagger}_{i\downarrow} c_{i\downarrow} - \frac{1}{2}\right)\,,
\end{align}
where the values $\varepsilon_i$ have been randomly drawn from a normal distribution of width $\sigma = t / 2$.
The corresponding Hamiltonian after Schrieffer-Wolff transformation and projection to the subspace
of the Hilbert space where $\sum_{\sigma} c^{\dagger}_{i\sigma} c_{i\sigma} = 1$ for every site $i$
of the chain reads
\begin{align}
    \mathcal{P} \tilde{H} \mathcal{P} = \sum_{i=1}^{N} J_i\, \vec{S}_i \cdot
    \vec{S}_{i-1} + \sum_{i=1}^{N} \varepsilon_i\,,
    \label{eq:h_xxx}
\end{align}
where $\vec{S}_{i}=(S^x_i, S^y_i, S^z_i)$ denotes the spin operator acting on the electron spin of the
electron located on chain site $i$ and the coupling constants $J_i = \frac{4t^2}{U} +\delta_i$ feature
small local renormalizations $\delta_i$ caused by the disorder.

\begin{figure}
    \begin{center}
        \includegraphics[width=0.49\textwidth]{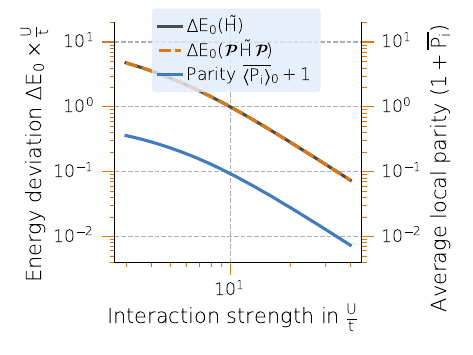}
        \caption{Rescaled energy difference $U\times\Delta E_0/t$ between the true ground state of the
            disordered Fermi-Hubbard chain and the ground states of $\tilde{H}$ (dark grey)
            and of the Heisenberg Hamiltonian $\mathcal{P} \tilde{H} \mathcal{P}$ (dashed orange), where the
chain sites are replaced with local spin degrees of freedom, and the true ground state. The blue line
displays the average local parity $\overline{\langle P_i \rangle}_0 = \frac{1}{N}\sum_{i=1}^{N}
\langle P_i \rangle_0$
of the individual chain sites $i$ in the ground state.}
        \label{fig:hubbard_energy}
    \end{center}
\end{figure}

We have examined a disordered Fermi-Hubbard Hamiltonian of length $N=10$ chain sites and periodic
boundary conditions $c_{i=0,\sigma} = c_{i=N,\sigma}$. The
hybridization $t$ between chain sites represents the energy scale of the system. We have
fixed the electron number $N_{\mathrm{electrons}}=10$ and $\sum_i \sigma^z_i=0$ to perform a ground state
computation using a numerical diagonalization method for a range of repulsive Hubbard interaction strength values $3 \leq U/t \leq 40$.
We have computed the average local parity $(\overline{\langle P_i \rangle}_0 + 1)$ of the
chain sites in the ground state. We have used our Schrieffer-Wolff transformation approach to
numerically determine the Hamiltonians $\tilde{H}$ and $\mathcal{P} \tilde{H} \mathcal{P}$ for each
value of the interaction strength $U/t$ and have compared them with the established
result~(\ref{eq:h_xxx}). The numerical results match the
analytic expressions exactly. We have again calculated the discrepancy $\Delta E_0$ between the ground state
energies $E_0$ of the dFH Hamiltonian $H$~(\ref{eq:h_dis_hubbard}) and the energy of respective ground states of the
transformed Hamiltonians $\tilde{H}$ and $\mathcal{P}\tilde{H}\mathcal{P}$. The ground state
average local
parity $\overline{\langle P_i \rangle_0} + 1$ of the chain sites and the rescaled ground energy
differences $U\Delta E_0 /t$ as a function of the interaction strength $U/t$ are displayed in
Fig.~\ref{fig:hubbard_energy}.

For values of the interaction strength $U/t \leq 10$, we
observe a significant energy difference $\Delta E_0$ for both $\tilde{H}$ and $\mathcal{P} \tilde{H}
\mathcal{P}$ which is decreasing with $U/t$ at an increasing rate
faster than $(U/t)^{-1}$. At values $U/t > 10$ the energy difference becomes small with respect
to the energy scale $t$ and is approaching $\Delta E_0 \rightarrow 0$ for $U/t \rightarrow \infty$
at a rate faster than $(U/t)^{-2}$. Similarly we find average local parity $(\overline{\langle P_i
\rangle_0} + 1) > 10^{-1}$ for $U/t \leq 10$. This highlights that the chain sites cannot be considered good
realizations of spin degrees of freedom in this regime. The transformed Hamiltonian $\tilde{H}$
can consequently not be considered a good approximation to the Hamiltonian $H$~(\ref{eq:h_dis_hubbard}).
This is supported by the respective values of $\Delta E_0/t \geq 1$. As $(\overline{\langle P_i
\rangle_0} + 1)(U/t) \ll 10^{-1}$ for $U/t > 10$ we also observe that the transformed Hamiltonian
becomes an ever better approximation of the original dFH Hamiltonian and $\Delta E_0/t \ll 1$.
For the disordered Fermi-Hubbard Hamiltonian we again find that our Schrieffer-Wolff transformation
method recovers the analytically known results.
The ground state of effective Hamiltonian $\mathcal{P} \tilde{H} \mathcal{P}$ approaches the
true ground state as $U/t \rightarrow \infty$ at a rate faster than $(U/t)^{-1}$ and the crossover
at which $\mathcal{P} \tilde{H} \mathcal{P}$ becomes a good description of the low-energy
dynamics of the model is characterized by $(\overline{\langle P_i
\rangle_0} + 1) < 10^{-1}$.
We note that the Schrieffer Wolff transformation renormalizes the local disorder potentials.

\subsection{Molecular Chromium bromide}
\label{sec:crbr}
An interesting testbed for our spin mapping approach, is the class of 2D magnetic materials, which have recently been
discovered experimentally\cite{Gong2017,Huang2017}--seemingly in contradiction to the Mermin-Wagner theorem. A common approach
to study 2D magnetic materials from first principles is to perform a so-called ``DFT+U'' calculation, followed by a Wannierization
and the extraction of a \emph{classical} Heisenberg-like spin model from the resulting tight-binding description of the material (cf.\
\cite{Esteras2023} for a recent study on Chromium tri-halides). The spin mapping proposed in this work represents an alternative for extracting
the effective spin model. The key difference to the aforementioned workflow to extract effective spin models using DFT is:
1) A correlated quantum-chemistry (wave-function) approach is used to determine a basis set featuring spin-like orbitals.
2) The Schrieffer--Wolff transformation starts from full description not relying on the approximate multi-reference method
   used to determine the spin-like orbitals.
3) The Hamiltonian resulting from the Schrieffer-Wolff transformation is still a quantum
mechanical Hamiltonian, i.e., the classical limit is not implied by the spin mapping approach.

Let us point out that we neither claim that a full
quantum mechanical treatment of the effective spin model is required for getting an accurate spin model for (ferromagnetic) transition-metal tri-halides,
nor that a correlated method is required for the electronic-structure calculation (however, standard DFT, without ``+U'', is not sufficient to
capture the localized Chromium $d$-orbitals).
We would also like to note that the main purpose of the multi reference calculation is the determination of the spin orbitals,
i.e.\ the single particle basis. We can use hints from multi reference methods to restrict the application of the Schrieffer-Wolff transformation
to a subset of all orbitals, but this step is not essential.

As a minimal model for the $\mathrm{Cr}\mathrm{Br}_3$ solid, we consider a minimal cluster, comprised of a single
Chromium atom, cut out from the periodic structure, together with an octahedron of six Bromide atoms coordinating the transition metal.
To study the spin mapping approach for inter-Chromium couplings larger clusters must be considered, which is beyond the scope of the present work.
As a first step towards studying chromium bromide $\mathrm{Cr}\mathrm{Br}_3$ we hence study a $\mathrm{Cr}\mathrm{Br}_6^{3-}$ cluster to test
the applicability of our approach.

\begin{figure}
    \begin{center}
        \includegraphics[width=0.49\textwidth]{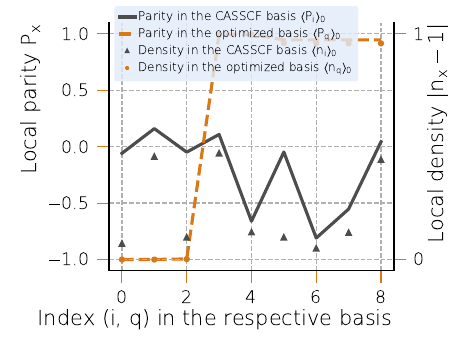}
        \caption{Local parities of the active space basis orbitals of molecular chromium bromide
            $\mathrm{Cr}\mathrm{Br}^{3-}_6$ in the CASSCF
            ground state. The dark grey line displays the CASSCF ground state local parities of the
            CASSCF canonical molecular orbitals, which form the active space of the system. The dashed orange line
            indicates the ground state local parities of the optimized basis orbitals $\phi_q$ and the orange circles show the
            expectation value for the local electrons density of basis orbitals $\phi_q$ in the ground state.
            The dark gray triangles indicate the expectation values for local electron density in the original CASSCF basis orbitals.
            In the optimized basis there exist three basis orbitals that feature local parities $\langle
            P_q\rangle_0 < -0.994$, indicating good realizations of three spin degrees of freedom.}
        \label{fig:crbr_parity}
    \end{center}
\end{figure}

We initially perform a series of electronic structure calculations using the def2-QZVP basis set to generate the information
necessary for a search of the active space of orbital basis of the system. We have used the Active Space Finder (ASF),~\cite{asf2024} an open-source
to assist the user in the selection of an active space. As expected, five Chromium $d$-orbitals and $p$-orbitals from the outer Shell
of Bromide are suggested by the ASF and the active space contains $N_{\phi_{i}}=9$ basis orbitals. We have performed a CASSCF calculation
using the previously identified orbitals as the active space to
compute the reduced density matrices $\rho^{(1)}$ and $\rho^{(2)}$ for the CASSCF ground state.
From this CASSCF calculation we have also obtained the one-electron $t^{\sigma\sigma'}_{ij}$ and
two-electron integrals $V^{\sigma\sigma'}_{ijkl}$ in the active space. We can
write the Hamiltonian description of the active space of chromium bromide as
\begin{align}
    H = \sum_{ij\sigma\sigma'} t^{\sigma \sigma'}_{ij} c^{\dagger}_{i\sigma} c_{j\sigma'} +
    \sum_{ijkl\sigma\sigma'} V^{\sigma\sigma'}_{ijkl} c^{\dagger}_{i\sigma} c^{\dagger}_{j\sigma'}
    c_{k\sigma'} c_{l\sigma}\,,
\end{align}
where $c^{\dagger}_{i\sigma}$ creates an electron with spin $\sigma$ in the orbital $\phi_i$ of the
active space.
Using the reduced density matrices we have performed a parity optimization of the basis orbitals of
the active space. Here, we first ran a parity minimization procedure of the entire basis and
identified $N_{\phi_q}=3$ spin-like orbital with respective parities $\langle P_q \rangle_0 \leq
-9.94\times 10^{-1}$. In a subsequent optimization run have maximized the local parities of the remaining basis
orbitals. The local parities of the initial basis orbitals $\phi_i$ and the
optimized basis orbitals $\phi_q$ are shown together with the average electron density $\langle n_q
\rangle_0$ in Fig.~\ref{fig:crbr_parity}.

We find that three orbitals $\phi_i$ of the original basis
already have smaller local
parities $\langle P_i \rangle_0 \leq -5.53\times 10^{-1}$. The spin-like orbitals $\phi_q$ of the
optimized basis feature significant contributions by
these initial basis orbitals $i=\lbrace 4, 6, 7\rbrace$. The orbitals $\phi_q$ read
\begin{align}
    \phi_{q=0} &= -0.81\, \phi_{7} - 0.49\, \phi_{8} + 0.20\, \phi_{5} + \dots\,, \\\nonumber
    \phi_{q=1} &= -0.86\, \phi_{4} - 0.20 \, \phi_{7} + \dots\,, \\\nonumber
    \phi_{q=2} &= +0.95\, \phi_{6} - 0.26\, \phi_{1} + \dots \,,
    \label{}
\end{align}
where $\dots$ denotes minor contributions from the other basis orbitals $\phi_i$.

We have performed a
first numerical diagonalization computation of the $n=20$ lowest energy eigenstates of the effective Hamiltonian $\mathcal{P}_0 H
\mathcal{P}_0$,
which acts exclusively on the subspace of the Hilbert space characterized by $n_i \equiv 1$, for
$i\in \lbrace 4, 6, 7\rbrace$. In the calculation we have fixed $N_{\mathrm{electrons}}=9$ and $\sum_i \sigma^z_i = 3$.
Subsequently, the Hamiltonian $H$ was transformed to the parity-optimized orbital basis and a second
numerical diagonalization with the same constraints for electron number and total spin $\sigma^z$ was performed.
For each of the $n=20$ lowest energy eigenstates of the Hamiltonian $H$ we have computed the
local parity $\langle P_{q\in\mathcal{S}}\rangle_n$ of the orbitals initially identified as
spin-like $\phi_{q\in\mathcal{S}}$, where
$\mathcal{S}=\lbrace 0,1,2\rbrace$. We have projected the Hamiltonian $H$ in the parity-optimized basis to the subspace of the
Hilbert space where the number of electrons in the spin-like orbitals $\phi_{q\in\mathcal{S}}$ is fixed $n_{q\in \mathcal{S}}
\equiv 1$. Using the same numerical diagonalization method we have computed the $n=20$ lowest energy
eigenstates of this projected Hamiltonian $\mathcal{P} H \mathcal{P}$.
For the Schrieffer-Wolff transformation we have introduced a cutoff $c= 10^{-5}\,\mathrm{Hartree}$
for the constituent terms of the Hamiltonian. Hamiltonian terms of coupling constant $\vert
h\vert$ smaller than the cutoff
have been neglected in the calculation of the transformed Hamiltonian $\tilde{H}$ to limit the
memory consumption of the computation. The singular value decomposition in our
Schrieffer-Wolff transformation approach was performed using a Krylov subspace method with a
maximum dimension of the subspace $\mathrm{dim}(\mathcal{K})=4000$.
The size of the vector spaces $\mathcal{V}$ depends on the choice of the cutoff.
For chromium bromide we determine $\mathrm{dim}(\mathcal{V}_0)(c) = 8760$ for the initial vector
space and $\mathrm{dim}(\mathcal{V}_1)(c)=3.6\times 10^{6}$ for the target vector space. The
prevalent interaction terms in the transformed Hamiltonian read
\begin{align}
    \tilde{H} =& \sum_{q\in\mathcal{S},\sigma} t^{\sigma\sigma}_{qq} c^{\dagger}_{q\sigma}
    c_{q\sigma} + \sum_{qp\in\mathcal{S}} h_{qp}\,\vec{S}_{q}\cdot \vec{S}_{p}\\\nonumber
    &+\sum_{q\in\mathcal{S}, ij\in\bar{\mathcal{S}},\sigma\sigma'} h^{\sigma\sigma'}_{qqij}\,
    c^{\dagger}_{q\sigma} c_{q\sigma'}
    c^{\dagger}_{i\sigma'} c_{j\sigma} \\\nonumber
    &+\sum_{ij\in\bar{\mathcal{S}},\sigma\sigma'} t^{\sigma \sigma'}_{ij} c^{\dagger}_{i\sigma} c_{j\sigma'} +
    \sum_{ijkl\in\bar{\mathcal{S}},\sigma\sigma'} V^{\sigma\sigma'}_{ijkl}\, c^{\dagger}_{i\sigma} c^{\dagger}_{j\sigma'}
    c_{k\sigma'} c_{l\sigma}\\\nonumber
    &+\dots \,,
    \label{}
\end{align}
and amount to $N_{\vert h \vert > c} = 41239$ terms with a coupling constant $\vert h \vert > c$.
The projection of the
transformed Hamiltonian to the subspace of the Hilbert space, in which the spin-like orbitals
$\phi_{q\in\mathcal{S}}$ are
replaced with spin degrees of freedom, is given by
\begin{align}
    \mathcal{P}\tilde{H}\mathcal{P} =
    &\sum_{qp\in\mathcal{S}} h_{qp}\,\vec{S}_{q}\cdot \vec{S}_{p} + \sum_{q\in\mathcal{S},
    ij\in\bar{\mathcal{S}},\sigma\sigma'} h^{\sigma\sigma'}_{qij}\,
    \vec{S}_q \cdot
    c^{\dagger}_{i\sigma'} \vec{\sigma} c_{j\sigma} \\\nonumber
    &+\sum_{ij\in\bar{\mathcal{S}},\sigma\sigma'} t^{\sigma \sigma'}_{ij} c^{\dagger}_{i\sigma}
    c_{j\sigma'} \\\nonumber
    &+\sum_{ijkl\in\bar{\mathcal{S}},\sigma\sigma'} V^{\sigma\sigma'}_{ijkl} c^{\dagger}_{i\sigma} c^{\dagger}_{j\sigma'}
    c_{k\sigma'} c_{l\sigma}\\\nonumber
    &+\dots \,,
    \label{}
\end{align}
where $\vec{S}_q=(S^x_q, S^y_q, S^z_q)$ and $\vec{\sigma}=(\sigma^x_{\sigma'\sigma},
\sigma^y_{\sigma'\sigma}, \sigma^z_{\sigma'\sigma})$.

\begin{figure}
    \begin{center}
        \includegraphics[width=0.49\textwidth]{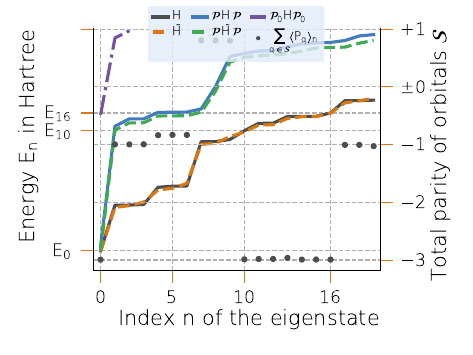}
        \caption{Low energy spectra of the active space Hamiltonians of molecular chromium bromide
            $\mathrm{Cr}\mathrm{Br}^{3-}_6$.
            The dark grey line shows the eigenenergies of the $n=20$ lowest eigenstates of the original
            active space Hamiltonian $H$. We display the corresponding eigenenergies of the effective Hamiltonian
            $\mathcal{P} H \mathcal{P}$ (solid blue), projected onto the subspace of the Hilbert space where the spin-like
            orbitals are restricted to single occupancy,
            the transformed Hamiltonian $\tilde{H}$ (dashed orange), and the effective
            Hamiltonian $\mathcal{P} \tilde{H} \mathcal{P}$ (dashed green) acting on the subspace $\mathcal{P}$ in which the
            fermionic degrees of freedom in
            the spin-like orbitals have been substituted with spin degree of freedom. The low energy spectrum of
            the effective Hamiltonian
            $\mathcal{P}_0 H \mathcal{P}_0$, acting only on the subspace characterized by $n_i \equiv 1$ for
            the CASSCF canonical orbitals $i\in\lbrace 4,6,7\rbrace$, is shown in purple.
            The total local parity
            of the spin-like orbitals $\sum_{q\in\mathcal{S}}\langle P_{q} \rangle_n$ measured for the lowest $n=20$ eigenstates
            $\vert n \rangle$ of the original fermionic Hamiltonian $H$ is displayed as grey circles.}
        \label{fig:crbr_spectrum}
    \end{center}
\end{figure}

We have calculated the low energy spectrum of $\tilde{H}$ and $\mathcal{P} \tilde{H} \mathcal{P}$
using a numerical diagonalization method and with the same set of constraints for the quantum numbers. The low-energy
spectra from numerical diagonalization calculations of the five different Hamiltonian representations are
displayed in Fig.~\ref{fig:crbr_spectrum} (a).
\begin{figure}
    \begin{center}
        \begin{subfigure}{0.49\textwidth}
            \centering
            \includegraphics[width=\textwidth]{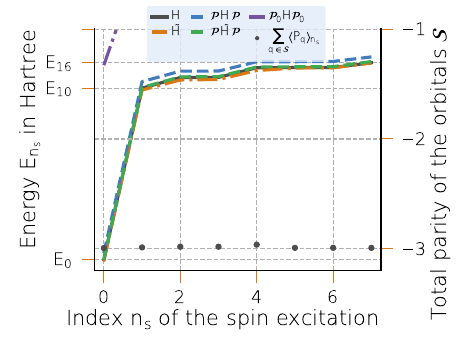}
            \caption{}
        \end{subfigure}
        \begin{subfigure}{0.49\textwidth}
            \centering
            \includegraphics[width=\textwidth]{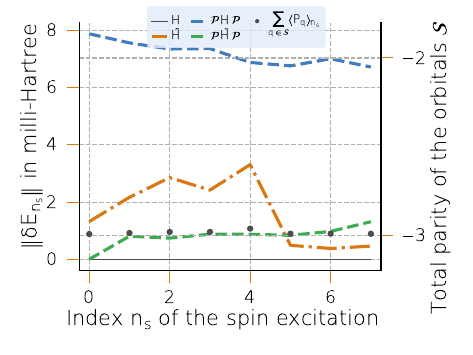}
            \caption{}
        \end{subfigure}
        \caption{(a) Low energy
spectra of the respective Hamiltonians reduced to the set of eigenstates $\vert
n_{\mathcal{S}}\rangle$ of $H$ and $\tilde{H}$, for which the expectation value of the total local
parity for the spin-like orbitals $\phi_q$ satisfies
$\sum_{q\in\mathcal{S}}\langle P_{q} \rangle_{n_{\mathcal{S}}} \simeq -3$. (b) Energy deviation
$\vert \delta E_{n_{\mathcal{S}}} \vert$
between the eigenergies $E_{n_{\mathcal{S}}}$ of the full fermionic Hamiltonian and the respective transformed and
projected, effective Hamiltonians. We observe that the spectrum of spinful excitations is better captured by
the transformed Hamiltonians $\mathcal{P}\tilde{H}\mathcal{P}$ as opposed to the original Hamiltonian projected onto the subspace of
the Hilbert space, where the spin-like orbitals become spin degrees of freedom,
$\mathcal{P}H\mathcal{P}$. The energy
discrepancy between $H$ and $\mathcal{P}\tilde{H}\mathcal{P}$ remains $\vert \delta
E_{n_{\mathcal{S}}}\vert \approx
1\times 10^{-3}\,\mathrm{Hartree}$ for the studied excited states. The corresponding discrepancy for
the effective Hamiltonian $\mathcal{P}H\mathcal{P}$ is noticeably larger and amounts to $\vert \delta
E_{n_{\mathcal{S}}}\vert \leq
8\times 10^{-3}\,\mathrm{Hartree}$.}
    \label{fig:crbr_energy}
    \end{center}
\end{figure}

We observe that the ground state energies of
the effective Hamiltonians $\mathcal{P} H \mathcal{P}$, $\tilde{H}$,
and $\mathcal{P}\tilde{H}\mathcal{P}$ are within a fraction of a Hartree from the true ground state
energy $E_{0}^{H}$
of the original fermionic Hamiltonian $H$. It is noteworthy that the ground state energy
$E_0^{\mathcal{P}\tilde{H}\mathcal{P}}$ of the
effective spin-bath Hamiltonian $\mathcal{P}\tilde{H}\mathcal{P}$ deviates from the true ground
state energy only on the order of $\mu\mathrm{Hartree}$.
The respective energy differences amount to
\begin{align*}
    \Delta E_{0}^{\mathcal{P}\tilde{H}\mathcal{P}}
     =& \,\left(E_{0}^{H} - E_{0}^{\mathcal{P}\tilde{H}\mathcal{P}}\right) \simeq 2\times 10^{-5}\,\mathrm{Hartree}
  \\ <& \,\left(E_{0}^{H} - E_{0}^{\tilde{H}}\right) \simeq 2\times 10^{-3}\,\mathrm{Hartree}
  \\ <& \,\left(E_{0}^{H} - E_{0}^{\mathcal{P}H\mathcal{P}}\right) \simeq 8\times 10^{-3}\,\mathrm{Hartree}\,.
\end{align*}
In contrast, we find a significantly larger energy discrepancy $\Delta E_0^{\mathcal{P}_0 H
\mathcal{P}_0} = (E_{0}^{H} - E_{0}^{\mathcal{P}_0 H\mathcal{P}_0}) \simeq 2\times
10^{-1}\,\mathrm{Hartree}$ for the naive effective spin-bath Hamiltonian $\mathcal{P}_0 H
\mathcal{P}_0$.

At first glance, we observe significant energy differences
between the original fermionic Hamiltonian $H$ and the effective spin-bath Hamiltonians $\mathcal{P} H \mathcal{P}$ and
$\mathcal{P} \tilde{H} \mathcal{P}$ for the excited states, while the transformed fermionic Hamiltonian $\tilde{H}$
appears to remain a reasonable approximation to the original Hamiltonian $H$ with $\vert \delta E_n
\vert =
\vert E_{n}^{H} - E_{n}
^{\tilde{H}}\vert \leq 3\times 10^{-3}\,\mathrm{Hartree}$. We find that the $n=9$ lowest excited
states correspond to an average local parity
$\sum_{q\in \mathcal{S}} \langle P_q \rangle_n \geq -1$. This implies the
presence of either no spin degree freedom or at most one distinct
local spin degree of freedom instead of a total of three distinct ones, such that the effective three-spin-system Hamiltonians $\mathcal{P} H \mathcal{P}$ and $\mathcal{P}
\tilde{H}\mathcal{P}$ are not sensible descriptions of the physics of the excited states with significant charge fluctuations. If we
reduce the spectrum to eigenstates of $H$ with three spin-like orbitals, i.e. total local parity
$\sum_{q\in{\mathcal{S}}}\langle P_{q} \rangle_{n_{\mathcal{S}}} \simeq -3$, a different picture
emerges. This reduced spectrum is shown in Fig.~\ref{fig:crbr_energy}~(a). The corresponding energy
discrepancies $\vert \delta E_{n_{\mathcal{S}}} \vert$ from the true eigen energies of $H$ are shown in Fig.~\ref{fig:crbr_energy}~(b). The effective Hamiltonians
$\mathcal{P} H \mathcal{P}$ and $\mathcal{P} \tilde{H} \mathcal{P}$ appear to capture this reduced spectrum
well. We observe a maximum energy discrepancy $\vert \delta E_{n_{\mathcal{S}}}\vert = 1.5\times
10^{-3}\,\mathrm{Hartree}$ for the effective transformed Hamiltonian $\mathcal{P}\tilde{H}\mathcal{P}$, while we
see a larger discrepancy $\vert \delta E_{n_{\mathcal{S}}}\vert = 8\times 10^{-3}\,\mathrm{Hartree}$ for the
effective Hamiltonian $\mathcal{P}H\mathcal{P}$.
The effective Hamiltonian $\mathcal{P}_0 H \mathcal{P}_0$ remains a bad approximation throughout this
reduced energy spectrum. This highlights the necessity of a suitable orbital basis for the
derivation of effective Hamiltonians describing the spin physics. In the
parity-optimized orbital basis, for the given example of
$\mathrm{Cr}\mathrm{Br}^{3-}_6$, the terms coupling to the charge degree of freedom of the electron
densities in the spin-like orbitals $\phi_{q\in\mathcal{S}}$ are suppressed to the extent
that the simple effective Hamiltonian $\mathcal{P} H \mathcal{P}$
already becomes a relatively good description of the spin excitation spectrum.
The subsequent Schrieffer-Wolff transformation and utilization of the corresponding effective
transformed Hamiltonian
$\mathcal{P} \tilde{H} \mathcal{P}$ then yield smaller additional improvements. This emphasizes the importance
of using a suitable single particle basis.
In constrast, we observe no such drastic
suppression of the undesirable coupling terms for the basis of the CASSCF canonical molecular
orbitals.

We have shown that our optimization procedure is able to correctly identify the $N_{\phi_q}=3$ spin-like
orbitals such that a subset of the energy spectrum of $\mathrm{Cr}\mathrm{Br}^{3-}_6$ can be accurately
reproduced by a system of three spin degrees of freedom coupled to a
fermionic environment. We further observe that our Schrieffer-Wolff transformation approach yields a
transformed Hamiltonian $\tilde{H}$ that, if projected to the Hilbert space representing three
spins in a fermionic environment, improves upon the description given by
the original Hamiltonian $H$ projected to the same subspace.
As shown in Fig.~\ref{fig:crbr_energy} the energy of the ground state and the low energy spin excited states are much closer
to the original model if we apply our Schrieffer-Wolff method before projecting the spin-like orbitals on pure spin orbitals.
We validate that the local parity of the spin-like
basis orbitals can be regarded as a good indicator of the accuracy of the effective spin-bath Hamiltonian description
$\mathcal{P}\tilde{H}\mathcal{P}$ derived with our method.

\section{Conclusion}
Effective model Hamiltonians provide enormous value for the investigation of individual aspects of
complex systems. Here, we have
presented a procedure for the automatic derivation of model Hamiltonians of coupled spin degrees of
freedom embedded in a fermionic environment. These Hamiltonians provide a means to
enhance the study of the dominant low temperature spin physics of complex real materials.
We have shown that to investigate the spin physics of such materials properly, one first needs to
operate within a suitable single particle orbital basis, where a subset of the basis orbitals become
realizations of spin degrees of freedom at low energies.
We have introduced the local parity as a sensible metric for the spin-like behavior of electron
densities in the basis orbitals. We have proven that the natural orbital basis
does not generally correspond to the basis that features the spin-like orbitals. We have provided an
example toy model
for which the natural orbital basis is significantly worse than the initial basis for the study of spin
physics. We also studied a well-known
singlet-diradical molecule, namely closed configuration para-benzyne, and demonstrated that the two
well-known non-trivial spin-like orbitals are correctly determined by our local parity optimization
procedure. We find that the optimization of the local parity yields an
orbital basis that typically results in a many particle basis of the Hilbert space with small
coupling and significant energy gap between the different subspaces.
Our extended Schrieffer-Wolff transformation method allows to also transform generic Hamiltonians with
complicated interaction terms beyond the density-density interactions of the toy models that the
Schrieffer-Wolff transformation has so far mostly been applied to.
There is also no need for the block-diagonal part $H_0$ to be easily diagonalizable, because no calculation of the spectrum
of $H_0$ is required in Schrieffer-Wolff transformation our method. We have shown that our Schrieffer-Wolff transformation method
reproduces established results for the well-known single impurity Anderson model and the disordered
Fermi-Hubbard model Hamiltonians. For these models we corroborate that the local parity of the spin-like
orbitals is a good predictor for the quality of the effective spin-bath Hamiltonian as the
description of the low energy spin dynamics. Lastly, we have applied the full spin-bath Hamiltonian
derivation procedure to molecular chromium bromide.
Here, we have shown that the derived effective spin-bath models are good descriptions of the system's low-energy
spin excitations.
We would like to point out that while we used an approximate multi reference method to determine the spin-like orbitals,
the Schrieffer--Wolff mapping bypasses the standard quantum chemistry methods directly projecting the quantum chemistry model on the
spin / spin-bath model. Being able to perform these calculations for quantum chemistry descriptions including all four index terms
is provided by our representation of the Schrieffer-Wolff equation as a linear set of equations.
Future work will address more complex materials, where the Schrieffer-Wolff
transformation becomes essential to obtain good effective Hamiltonian descriptions for the
low-energy spin dynamics of the material.

\section*{Data availability}

The numerical data can be obtained from the authors on request. The spin finding procedure and the imple-
mented Schrieffer Wolff transformation can be tested on https://cloud.quantumsimulations.de/ using the HQS
Spin Mapper module.

\begin{acknowledgments}
    We thank Reza Shirazi for insightful discussions.
    BMS, NE, and PS were supported by the Federal Ministry of Education and Research (BMBF) through the QSolid
    project (Grant no. 13N16155). BMS and PS acknowledge support by the BMBF through the MANIQU project
    (Grant no. 13N15576).
    VR was supported by the Federal Ministry for Economic Affairs and Climate Action through the AQUAS project (Grant no. 01MQ22003A).
\end{acknowledgments}

\clearpage
\appendix
\section{Derivatives of the local parity}
The analytic expression for the local parity $\langle P_q \rangle$ of the orbital $\phi_q$ after rotation as a function of
the rotation angle $\theta$ and its derivatives are given by
\label{sec:diff_p}
\begin{widetext}
\begin{align}
    \langle P_{q} \rangle =& - 2\left[\sum_{\sigma}\cos^2 \theta\, \langle c^{\dagger}_{i\sigma}
    c_{i\sigma}\rangle + \sin^2 \theta \, \langle c^{\dagger}_{j\sigma} c_{j\sigma}\rangle + \cos
\theta \sin \theta (\langle c^{\dagger}_{i\sigma} c_{j\sigma}\rangle + \langle c^{\dagger}_{j\sigma} c_{i\sigma}\rangle) \right]\nonumber\\
    &+4 \left[ \cos^4 \theta \, \langle c^{\dagger}_{i\uparrow} c^{\dagger}_{i\downarrow}
    c_{i\downarrow} c_{i\uparrow} \rangle + \sin^4 \theta \, \langle c^{\dagger}_{j\uparrow} c^{\dagger}_{j\downarrow} c_{j\downarrow} c_{j\uparrow}\rangle \right.\nonumber\\
    &\qquad + \cos^3 \theta \sin \theta \left(\langle c^{\dagger}_{i\uparrow} c^{\dagger}_{i\downarrow} c_{i\downarrow} c_{j\uparrow} \rangle + \langle c^{\dagger}_{i\uparrow} c^{\dagger}_{i\downarrow} c_{j\downarrow} c_{i\uparrow} \rangle + \langle c^{\dagger}_{i\uparrow} c^{\dagger}_{j\downarrow} c_{i\downarrow} c_{i\uparrow} \rangle + \langle c^{\dagger}_{j\uparrow} c^{\dagger}_{i\downarrow} c_{i\downarrow} c_{i\uparrow} \rangle \right)\nonumber\\
    &\qquad + \cos \theta \sin^3 \theta \left(\langle c^{\dagger}_{j\uparrow} c^{\dagger}_{j\downarrow} c_{j\downarrow} c_{i\uparrow} \rangle + \langle c^{\dagger}_{j\uparrow} c^{\dagger}_{j\downarrow} c_{i\downarrow} c_{j\uparrow} \rangle + \langle c^{\dagger}_{j\uparrow} c^{\dagger}_{i\downarrow} c_{j\downarrow} c_{j\uparrow}\rangle + \langle c^{\dagger}_{i\uparrow} c^{\dagger}_{j\downarrow} c_{j\downarrow} c_{j\uparrow}\rangle \right)\nonumber\\
    &\qquad + \cos^2 \theta \sin^2 \theta \left(\langle c^{\dagger}_{i\uparrow} c^{\dagger}_{i\downarrow} c_{j\downarrow} c_{j\uparrow}\rangle + \langle c^{\dagger}_{i\uparrow} c^{\dagger}_{j\downarrow} c_{j\downarrow} c_{i\uparrow} \rangle + \langle c^{\dagger}_{j\uparrow} c^{\dagger}_{j\downarrow} c_{i\downarrow} c_{i\uparrow}\rangle + \langle c^{\dagger}_{j\uparrow} c^{\dagger}_{i\downarrow} c_{i\downarrow} c_{j\uparrow}\rangle \right.\nonumber\\
    &\qquad \left. \left.+ \langle c^{\dagger}_{j\uparrow} c^{\dagger}_{i\downarrow} c_{j\downarrow} c_{i\uparrow}\rangle + \langle c^{\dagger}_{i\uparrow} c^{\dagger}_{j\downarrow} c_{i\downarrow} c_{j\uparrow}\rangle \right)\right] + 1 \,,
\end{align}
\end{widetext}
\begin{widetext}
\begin{align}
    \frac{d\langle P_{q} \rangle}{d\theta} =& 4 \sin \theta \cos \theta \left(\sum_{\sigma} \langle
    c^{\dagger}_{i\sigma} c_{i\sigma} \rangle - \langle c^{\dagger}_{j\sigma} c_{j\sigma}
\rangle\right)+2 (\sin ^2\theta -  \cos ^2 \theta)\left(\sum_{\sigma} \langle c^{\dagger}_{i\sigma} c_{j\sigma}\rangle + \langle c^{\dagger}_{j\sigma} c_{i\sigma}\rangle\right) \nonumber\\
     &+4 \left[(-4) \sin \theta \cos^3 \theta \,\langle c^{\dagger}_{i\uparrow}
     c^{\dagger}_{i\downarrow} c_{i\downarrow} c_{i\uparrow}\rangle + 4 \sin ^3 \theta \cos \theta\, \langle c^{\dagger}_{j\uparrow} c^{\dagger}_{j\downarrow} c_{j\downarrow} c_{j\uparrow}\rangle \right.\nonumber\\
     &\qquad +2 \left(\sin \theta \cos ^3 \theta - \sin ^3 \theta \cos \theta \right) \langle A \rangle \nonumber \\
     &\qquad +\left(-\sin ^4 \theta +3 \sin ^2 \theta \cos ^2 \theta \right) \langle B \rangle \nonumber\\
     &\qquad \left. +\left(\cos ^4 \theta -3 \sin ^2 \theta \cos ^2 \theta\right) \langle C \rangle \right]\,,
\end{align}
\end{widetext}
\begin{widetext}
\begin{align}
    \frac{d^2 \langle P_{q} \rangle}{d\theta^2} =&\left[\sum_{\sigma} 4 \left(\cos ^2 \theta - \sin
        ^2 \theta \right) \left(\langle c^{\dagger}_{i\sigma} c_{i\sigma}\rangle - \langle c^{\dagger}_{j\sigma} c_{j\sigma} \rangle \right)
    +8 \sin \theta \cos \theta \left( \langle c^{\dagger}_{i\sigma} c_{j\sigma}\rangle + \langle c^{\dagger}_{j\sigma} c_{i\sigma}\rangle \right) \right]\nonumber\\
    &+4 \left[\left(-4  \cos ^4\theta +12 \sin ^2\theta \cos ^2\theta \right) \langle c^{\dagger}_{i\uparrow} c^{\dagger}_{i\downarrow} c_{i\downarrow} c_{i\uparrow}\rangle\right.\nonumber\\
    &\qquad +\left(-4\sin ^4\theta +12 \sin ^2\theta \cos ^2\theta\right)\langle c^{\dagger}_{j\uparrow} c^{\dagger}_{j\downarrow} c_{j\downarrow} c_{j\uparrow}\rangle\nonumber\\
    &\qquad +\left(2\sin ^4\theta +2 \cos ^4\theta -12 \sin ^2\theta \cos ^2\theta\right) \langle A \rangle\nonumber\\
    &\qquad +\left(6 \sin \theta \cos ^3\theta -10 \sin ^3\theta \cos \theta\right)\langle B \rangle \nonumber\\
    &\qquad\left. + \left(-10 \sin \theta \cos ^3\theta +6 \sin ^3\theta \cos \theta \right) \langle C \rangle \right]\,,
\end{align}
\end{widetext}
where we have abbreviated
\begin{widetext}
\begin{align}
    \langle A \rangle =& \left(\langle c^{\dagger}_{i\uparrow} c^{\dagger}_{i\downarrow}
    c_{j\downarrow} c_{j\uparrow} \rangle + \langle c^{\dagger}_{i\uparrow}
c^{\dagger}_{j\downarrow} c_{j\downarrow} c_{i\uparrow}\rangle + \langle c^{\dagger}_{j\uparrow}
c^{\dagger}_{j\downarrow} c_{i\downarrow} c_{i\uparrow}\rangle\right.\nonumber\\
     &\left. + \langle c^{\dagger}_{j\uparrow} c^{\dagger}_{i\downarrow} c_{i\downarrow} c_{j\uparrow}\rangle + \langle c^{\dagger}_{j\uparrow} c^{\dagger}_{i\downarrow} c_{j\downarrow} c_{i\uparrow}\rangle + \langle c^{\dagger}_{i\uparrow} c^{\dagger}_{j\downarrow} c_{i\downarrow} c_{j\uparrow}\rangle\right)\,, \nonumber\\
    \langle B \rangle =& \left(\langle c^{\dagger}_{j\uparrow} c^{\dagger}_{j\downarrow} c_{j\downarrow} c_{i\uparrow}\rangle + \langle c^{\dagger}_{j\uparrow} c^{\dagger}_{j\downarrow} c_{i\downarrow} c_{j\uparrow}\rangle + \langle c^{\dagger}_{j\uparrow} c^{\dagger}_{i\downarrow} c_{j\downarrow} c_{j\uparrow}\rangle + \langle c^{\dagger}_{i\uparrow} c^{\dagger}_{j\downarrow} c_{j\downarrow} c_{j\uparrow}\rangle \right)\,, \nonumber\\
    \langle C \rangle =& \left(\langle c^{\dagger}_{i\uparrow} c^{\dagger}_{i\downarrow} c_{i\downarrow} c_{j\uparrow}\rangle + \langle c^{\dagger}_{i\uparrow} c^{\dagger}_{i\downarrow} c_{j\downarrow} c_{i\uparrow}\rangle + \langle c^{\dagger}_{i\uparrow} c^{\dagger}_{j\downarrow} c_{i\downarrow} c_{i\uparrow}\rangle + \langle c^{\dagger}_{j\uparrow} c^{\dagger}_{i\downarrow} c_{i\downarrow} c_{i\uparrow}\rangle \right) \,. \nonumber
\end{align}
\end{widetext}
\section{Symmetry-specification of fermionic creation and annihilation operators}
\label{sec:sym_ops}
We demonstrate the concept of symmetry-specified block-offdiagonal operators for the case of fermionic creation and annihilation operators. For a single orbital the local Hilbert space is given by
\begin{align}
    \mathcal{H}_{\mathrm{local}} = \left\lbrace \vert 0 \rangle,\, \vert \uparrow \rangle,\, \vert \downarrow \rangle,\, \vert \uparrow\downarrow \rangle \right\rbrace\,,
\end{align}
where we have fixed the order $\uparrow,\, \downarrow$ for the fermionic sign. The fermionic annihilation operator $
c_{\downarrow}: \mathcal{H}_{\textrm{local}} \rightarrow \mathcal{H}_{\mathrm{local}}$
for fermions with spin $s_z =-\frac{1}{2}$ acts on the vectors in the local Hilbert space
and performs the linear map
\begin{align}
    &c_{\downarrow}(a_{0}\vert 0 \rangle + a_{\uparrow}\vert \uparrow \rangle + a_{\downarrow}\vert
    \downarrow \rangle + a_{\uparrow\downarrow} \vert \uparrow\downarrow \rangle)\\\nonumber
    &=a_0 \mathbf{0}_{\mathcal{H}} + a_{\uparrow} \mathbf{0}_{\mathcal{H}} + a_{\downarrow} \vert 0 \rangle + (-a_{\uparrow\downarrow})\vert \uparrow \rangle \,,
\end{align}
where $\mathbf{0}_{\mathcal{H}}$ denotes the zero element of the Hilbert space. We see that $c_{\downarrow}$ maps two of the states, namely $\vert 0 \rangle$ and $\vert \uparrow \rangle$, map to the zero element.
There are thus two non-vanishing components of $c_{\downarrow}$ which map $\vert \downarrow \rangle \rightarrow \vert 0 \rangle$ and $\vert \uparrow\downarrow \rangle \rightarrow \vert \uparrow \rangle$.
We can choose to distinguish the blocks of the Hilbert space based on their total particle number $n = \sum_{\sigma} c^{\dagger}_{\sigma} c_{\sigma}$.
We consequently identify the state $\vert \uparrow \downarrow \rangle$ as belonging to the block characterized by the eigenvalue $\Lambda_q = n = 2$ and state $\vert \downarrow \rangle$ to the block characterized by the eigenvalue $\Lambda_q = n = 1$.
Given the choice of $n$ as the operator distinguishing the blocks of the Hilbert space, an intuitive
choice for the operator $A$ is the local number of fermions with opposite spin $s_z =
\frac{1}{2}$, i.e. the operator $n_{\uparrow} = c^{\dagger}_{\uparrow} c_{\uparrow}$. It is easily seen that the choice $A = n_{\uparrow}$ satisfies
\begin{align}
    \left[n,\, n_{\uparrow}\right] &= 0 \label{eq:block-dia}\,,\\
    \left[c_{\downarrow},\, n_{\uparrow}\right] &= 0\,,
\end{align}
where eq.~(\ref{eq:block-dia}) highlights that $n_{\uparrow}$ commutes with all operators that are
block-diagonal in our choice of blocks distinguish by $n$. The operator $n_{\uparrow}$ has
eigenvalues $\lambda_q \in \left\lbrace 0, 1 \right\rbrace$. Using
equations~(\ref{eq:operator_splitting}) and~(\ref{eq:splitting_prefactor}) we can express the operator $c_{\downarrow}$ as sum of unique components
\begin{align}
    c_{\downarrow} &= \alpha_{0}\left(n_{\uparrow} -1 \right)c_{\downarrow} + \alpha_{1}  \left(n_{\uparrow} -0\right)c_{\downarrow}\nonumber\\
    &= -(n_{\uparrow} - 1)c_{\downarrow} + (n_{\uparrow} - 0)c_{\downarrow}\nonumber\\
    &= (1 - n_{\uparrow})c_{\downarrow} + n_{\uparrow} c_{\downarrow} \nonumber\\
    &=c_{\lambda_q=0,\downarrow} + c_{\lambda_q=1,\downarrow}\,.
\end{align}
The separation of $c_{\downarrow}$ into its unique components can also be achieved by multiplying it with the identity $\mathbf{1} = (1 - n_{\uparrow}) + n_{\uparrow}$.
We expand the Hermitian conjugate operator $c^{\dagger}_{\downarrow}$ in the same way as
\begin{align}
    c^{\dagger}_{\downarrow} = c^{\dagger}_{\lambda_q = 0, \downarrow} + c^{\dagger}_{\lambda_q = 1, \downarrow} =
    \left(1 - n_{\uparrow}\right) c^{\dagger}_{\downarrow} + n_{\uparrow} c^{\dagger}_{\downarrow}\,,
\end{align}
and find
\begin{align}
    \left[c_{\lambda_q=0, \downarrow},\, c^{\dagger}_{\lambda_q=0, \downarrow}\right] &=
    \left(1-n_{\uparrow}\right)\left(1-n_{\uparrow}\right) \left[c_{\downarrow},\,
    c^{\dagger}_{\downarrow}\right]\nonumber\\
    &= \left(1 - n_{\uparrow}\right)\left(1 - 2 c^{\dagger}_{\downarrow} c_{\downarrow}\right) \nonumber\\
    \left[c_{\lambda_q=0, \downarrow},\, c^{\dagger}_{\lambda_q=1, \downarrow}\right] &= \left(1-n_{\uparrow}\right) n_{\uparrow} \left[c_{\downarrow},\, c^{\dagger}_{\downarrow}\right] = 0 \nonumber\\
    \left[c_{\lambda_q=1, \downarrow},\, c^{\dagger}_{\lambda_q=0, \downarrow}\right] &= n_{\uparrow}\left(1-n_{\uparrow}\right) \left[c_{\downarrow},\, c^{\dagger}_{\downarrow}\right] = 0 \nonumber\\
    \left[c_{\lambda_q=1, \downarrow},\, c^{\dagger}_{\lambda_q=1, \downarrow}\right] &= n_{\uparrow}\, n_{\uparrow} \left[c_{\downarrow},\, c^{\dagger}_{\downarrow}\right] = n_{\uparrow} \left(1 - 2 c^{\dagger}_{\downarrow} c_{\downarrow}\right)\,.
\end{align}

\bibliographystyle{apsrev4-1}
\bibliography{References_SpinMapper.bib}

%merlin.mbs apsrev4-1.bst 2010-07-25 4.21a (PWD, AO, DPC) hacked
%Control: key (0)
%Control: author (72) initials jnrlst
%Control: editor formatted (1) identically to author
%Control: production of article title (-1) disabled
%Control: page (0) single
%Control: year (1) truncated
%Control: production of eprint (0) enabled
\begin{thebibliography}{49}%
\makeatletter
\providecommand \@ifxundefined [1]{%
 \@ifx{#1\undefined}
}%
\providecommand \@ifnum [1]{%
 \ifnum #1\expandafter \@firstoftwo
 \else \expandafter \@secondoftwo
 \fi
}%
\providecommand \@ifx [1]{%
 \ifx #1\expandafter \@firstoftwo
 \else \expandafter \@secondoftwo
 \fi
}%
\providecommand \natexlab [1]{#1}%
\providecommand \enquote  [1]{``#1''}%
\providecommand \bibnamefont  [1]{#1}%
\providecommand \bibfnamefont [1]{#1}%
\providecommand \citenamefont [1]{#1}%
\providecommand \href@noop [0]{\@secondoftwo}%
\providecommand \href [0]{\begingroup \@sanitize@url \@href}%
\providecommand \@href[1]{\@@startlink{#1}\@@href}%
\providecommand \@@href[1]{\endgroup#1\@@endlink}%
\providecommand \@sanitize@url [0]{\catcode `\\12\catcode `\$12\catcode
  `\&12\catcode `\#12\catcode `\^12\catcode `\_12\catcode `\%12\relax}%
\providecommand \@@startlink[1]{}%
\providecommand \@@endlink[0]{}%
\providecommand \url  [0]{\begingroup\@sanitize@url \@url }%
\providecommand \@url [1]{\endgroup\@href {#1}{\urlprefix }}%
\providecommand \urlprefix  [0]{URL }%
\providecommand \Eprint [0]{\href }%
\providecommand \doibase [0]{http://dx.doi.org/}%
\providecommand \selectlanguage [0]{\@gobble}%
\providecommand \bibinfo  [0]{\@secondoftwo}%
\providecommand \bibfield  [0]{\@secondoftwo}%
\providecommand \translation [1]{[#1]}%
\providecommand \BibitemOpen [0]{}%
\providecommand \bibitemStop [0]{}%
\providecommand \bibitemNoStop [0]{.\EOS\space}%
\providecommand \EOS [0]{\spacefactor3000\relax}%
\providecommand \BibitemShut  [1]{\csname bibitem#1\endcsname}%
\let\auto@bib@innerbib\@empty
%</preamble>
\bibitem [{\citenamefont {Heisenberg}(1928)}]{heisenberg_1928}%
  \BibitemOpen
  \bibfield  {author} {\bibinfo {author} {\bibfnamefont {W.}~\bibnamefont
  {Heisenberg}},\ }\href {\doibase 10.1007/BF01328601} {\bibfield  {journal}
  {\bibinfo  {journal} {Zeitschrift für Physik}\ }\textbf {\bibinfo {volume}
  {49}},\ \bibinfo {pages} {619–636} (\bibinfo {year} {1928})}\BibitemShut
  {NoStop}%
\bibitem [{\citenamefont {G\"unther}(2013)}]{Gunt2013}%
  \BibitemOpen
  \bibfield  {author} {\bibinfo {author} {\bibfnamefont {H.}~\bibnamefont
  {G\"unther}},\ }\href@noop {} {\emph {\bibinfo {title} {{NMR Spectroscopy:}
  {Basic Principles}, {Concepts} and {Applications} in {Chemistry}}}}\
  (\bibinfo  {publisher} {John Wiley \& Sons},\ \bibinfo {year}
  {2013})\BibitemShut {NoStop}%
\bibitem [{\citenamefont {Veryazov}\ \emph {et~al.}(2011)\citenamefont
  {Veryazov}, \citenamefont {Malmqvist},\ and\ \citenamefont
  {Roos}}]{Lund2011}%
  \BibitemOpen
  \bibfield  {author} {\bibinfo {author} {\bibfnamefont {V.}~\bibnamefont
  {Veryazov}}, \bibinfo {author} {\bibfnamefont {P.~{\AA}.}\ \bibnamefont
  {Malmqvist}}, \ and\ \bibinfo {author} {\bibfnamefont {B.~O.}\ \bibnamefont
  {Roos}},\ }\href {\doibase 10.1002/qua.23068} {\bibfield  {journal} {\bibinfo
   {journal} {International Journal of Quantum Chemistry}\ }\textbf {\bibinfo
  {volume} {111}},\ \bibinfo {pages} {3329–3338} (\bibinfo {year} {2011})},\
  \Eprint
  {http://arxiv.org/abs/https://onlinelibrary.wiley.com/doi/pdf/10.1002/qua.23068}
  {https://onlinelibrary.wiley.com/doi/pdf/10.1002/qua.23068} \BibitemShut
  {NoStop}%
\bibitem [{\citenamefont {McArdle}\ \emph {et~al.}(2018)\citenamefont
  {McArdle}, \citenamefont {Endo}, \citenamefont {Aspuru-Guzik}, \citenamefont
  {Benjamin},\ and\ \citenamefont {Yuan}}]{McAr2018}%
  \BibitemOpen
  \bibfield  {author} {\bibinfo {author} {\bibfnamefont {S.}~\bibnamefont
  {McArdle}}, \bibinfo {author} {\bibfnamefont {S.}~\bibnamefont {Endo}},
  \bibinfo {author} {\bibfnamefont {A.}~\bibnamefont {Aspuru-Guzik}}, \bibinfo
  {author} {\bibfnamefont {S.}~\bibnamefont {Benjamin}}, \ and\ \bibinfo
  {author} {\bibfnamefont {X.}~\bibnamefont {Yuan}},\ }\href
  {http://arxiv.org/abs/1808.10402} {\bibfield  {journal} {\bibinfo  {journal}
  {arXiv:1808.10402 [quant-ph]}\ } (\bibinfo {year} {2018})},\ \bibinfo {note}
  {arXiv: 1808.10402}\BibitemShut {NoStop}%
\bibitem [{\citenamefont {Pouse}\ \emph {et~al.}(2023)\citenamefont {Pouse},
  \citenamefont {Peeters}, \citenamefont {Hsueh}, \citenamefont {Gennser},
  \citenamefont {Cavanna}, \citenamefont {Kastner}, , \citenamefont
  {Mitchell},\ and\ \citenamefont {Goldhaber-Gordon}}]{Pouse2023}%
  \BibitemOpen
  \bibfield  {author} {\bibinfo {author} {\bibfnamefont {W.}~\bibnamefont
  {Pouse}}, \bibinfo {author} {\bibfnamefont {L.}~\bibnamefont {Peeters}},
  \bibinfo {author} {\bibfnamefont {C.}~\bibnamefont {Hsueh}}, \bibinfo
  {author} {\bibfnamefont {U.}~\bibnamefont {Gennser}}, \bibinfo {author}
  {\bibfnamefont {A.}~\bibnamefont {Cavanna}}, \bibinfo {author} {\bibfnamefont
  {M.~A.}\ \bibnamefont {Kastner}}, , \bibinfo {author} {\bibfnamefont {A.~K.}\
  \bibnamefont {Mitchell}}, \ and\ \bibinfo {author} {\bibfnamefont
  {D.}~\bibnamefont {Goldhaber-Gordon}},\ }\href
  {"https://doi.org/10.1038/s41567-022-01905-4"} {\bibfield  {journal}
  {\bibinfo  {journal} {Nat. Phys.}\ ,\ \bibinfo {pages} {492–499}} (\bibinfo
  {year} {2023})}\BibitemShut {NoStop}%
\bibitem [{\citenamefont {Shirazi}\ \emph {et~al.}(2024)\citenamefont
  {Shirazi}, \citenamefont {Schoenauer}, \citenamefont {Schmitteckert},
  \citenamefont {Marthaler},\ and\ \citenamefont {Rybkin}}]{Shirazi2024}%
  \BibitemOpen
  \bibfield  {author} {\bibinfo {author} {\bibfnamefont {R.~G.}\ \bibnamefont
  {Shirazi}}, \bibinfo {author} {\bibfnamefont {B.~M.}\ \bibnamefont
  {Schoenauer}}, \bibinfo {author} {\bibfnamefont {P.}~\bibnamefont
  {Schmitteckert}}, \bibinfo {author} {\bibfnamefont {M.}~\bibnamefont
  {Marthaler}}, \ and\ \bibinfo {author} {\bibfnamefont {V.}~\bibnamefont
  {Rybkin}},\ }\href@noop {} {\enquote {\bibinfo {title} {Understanding
  radicals via orbital parities},}\ } (\bibinfo {year} {2024}),\ \Eprint
  {http://arxiv.org/abs/2404.18787} {arXiv:2404.18787 [physics.chem-ph]}
  \BibitemShut {NoStop}%
\bibitem [{\citenamefont {Schrieffer}\ and\ \citenamefont
  {Wolff}(1966)}]{Schr1966}%
  \BibitemOpen
  \bibfield  {author} {\bibinfo {author} {\bibfnamefont {J.}~\bibnamefont
  {Schrieffer}}\ and\ \bibinfo {author} {\bibfnamefont {P.}~\bibnamefont
  {Wolff}},\ }\href {\doibase 10.1103/PhysRev.149.491} {\bibfield  {journal}
  {\bibinfo  {journal} {Physical Review}\ }\textbf {\bibinfo {volume} {149}},\
  \bibinfo {pages} {491} (\bibinfo {year} {1966})}\BibitemShut {NoStop}%
\bibitem [{\citenamefont {Foldy}\ and\ \citenamefont
  {Wouthuysen}(1950)}]{Fold1950}%
  \BibitemOpen
  \bibfield  {author} {\bibinfo {author} {\bibfnamefont {L.~L.}\ \bibnamefont
  {Foldy}}\ and\ \bibinfo {author} {\bibfnamefont {S.~A.}\ \bibnamefont
  {Wouthuysen}},\ }\href {\doibase 10.1103/PhysRev.78.29} {\bibfield  {journal}
  {\bibinfo  {journal} {Physical Review}\ }\textbf {\bibinfo {volume} {78}},\
  \bibinfo {pages} {29} (\bibinfo {year} {1950})},\ \bibinfo {note} {publisher:
  American Physical Society}\BibitemShut {NoStop}%
\bibitem [{\citenamefont {Bravyi}\ \emph {et~al.}(2011)\citenamefont {Bravyi},
  \citenamefont {DiVincenzo},\ and\ \citenamefont {Loss}}]{Brav2011}%
  \BibitemOpen
  \bibfield  {author} {\bibinfo {author} {\bibfnamefont {S.}~\bibnamefont
  {Bravyi}}, \bibinfo {author} {\bibfnamefont {D.}~\bibnamefont {DiVincenzo}},
  \ and\ \bibinfo {author} {\bibfnamefont {D.}~\bibnamefont {Loss}},\ }\href
  {\doibase 10.1016/j.aop.2011.06.004} {\bibfield  {journal} {\bibinfo
  {journal} {Annals of Physics}\ }\textbf {\bibinfo {volume} {326}},\ \bibinfo
  {pages} {2793} (\bibinfo {year} {2011})},\ \bibinfo {note} {arXiv:
  1105.0675}\BibitemShut {NoStop}%
\bibitem [{\citenamefont {Landi}(2024)}]{Landi2024}%
  \BibitemOpen
  \bibfield  {author} {\bibinfo {author} {\bibfnamefont {G.~T.}\ \bibnamefont
  {Landi}},\ }\href {\doibase 10.48550/arXiv.2409.10656} {\enquote {\bibinfo
  {title} {Eigenoperator approach to {Schrieffer}-{Wolff} perturbation theory
  and dispersive interactions},}\ } (\bibinfo {year} {2024}),\ \bibinfo {note}
  {arXiv:2409.10656 [cond-mat, physics:hep-th, physics:quant-ph]}\BibitemShut
  {NoStop}%
\bibitem [{\citenamefont {Wurtz}\ \emph {et~al.}(2019)\citenamefont {Wurtz},
  \citenamefont {Claeys},\ and\ \citenamefont {Polkovnikov}}]{Wurt2019}%
  \BibitemOpen
  \bibfield  {author} {\bibinfo {author} {\bibfnamefont {J.}~\bibnamefont
  {Wurtz}}, \bibinfo {author} {\bibfnamefont {P.}~\bibnamefont {Claeys}}, \
  and\ \bibinfo {author} {\bibfnamefont {A.}~\bibnamefont {Polkovnikov}},\
  }\href {\doibase 10.1103/PhysRevB.101.014302} {\bibfield  {journal} {\bibinfo
   {journal} {arXiv:1910.11889 [cond-mat]}\ } (\bibinfo {year} {2019}),\
  10.1103/PhysRevB.101.014302},\ \bibinfo {note} {arXiv:
  1910.11889}\BibitemShut {NoStop}%
\bibitem [{\citenamefont {Wegner}(1994)}]{Wegn1994}%
  \BibitemOpen
  \bibfield  {author} {\bibinfo {author} {\bibfnamefont {F.}~\bibnamefont
  {Wegner}},\ }\href {\doibase 10.1002/andp.19945060203} {\bibfield  {journal}
  {\bibinfo  {journal} {Annalen der Physik}\ }\textbf {\bibinfo {volume}
  {506}},\ \bibinfo {pages} {77} (\bibinfo {year} {1994})}\BibitemShut
  {NoStop}%
\bibitem [{\citenamefont {Krull}\ \emph {et~al.}(2012)\citenamefont {Krull},
  \citenamefont {Drescher},\ and\ \citenamefont {Uhrig}}]{Krul2012}%
  \BibitemOpen
  \bibfield  {author} {\bibinfo {author} {\bibfnamefont {H.}~\bibnamefont
  {Krull}}, \bibinfo {author} {\bibfnamefont {N.~A.}\ \bibnamefont {Drescher}},
  \ and\ \bibinfo {author} {\bibfnamefont {G.~S.}\ \bibnamefont {Uhrig}},\
  }\href {\doibase 10.1103/PhysRevB.86.125113} {\bibfield  {journal} {\bibinfo
  {journal} {Physical Review B}\ }\textbf {\bibinfo {volume} {86}},\ \bibinfo
  {pages} {125113} (\bibinfo {year} {2012})},\ \bibinfo {note} {publisher:
  American Physical Society}\BibitemShut {NoStop}%
\bibitem [{\citenamefont {Schmiedinghoff}\ and\ \citenamefont
  {Uhrig}(2022)}]{Schm2022}%
  \BibitemOpen
  \bibfield  {author} {\bibinfo {author} {\bibfnamefont {G.}~\bibnamefont
  {Schmiedinghoff}}\ and\ \bibinfo {author} {\bibfnamefont {G.~S.}\
  \bibnamefont {Uhrig}},\ }\href {\doibase 10.21468/SciPostPhys.13.6.122}
  {\bibfield  {journal} {\bibinfo  {journal} {SciPost Physics}\ }\textbf
  {\bibinfo {volume} {13}},\ \bibinfo {pages} {122} (\bibinfo {year}
  {2022})}\BibitemShut {NoStop}%
\bibitem [{\citenamefont {White}(2002)}]{White:2002}%
  \BibitemOpen
  \bibfield  {author} {\bibinfo {author} {\bibfnamefont {S.~R.}\ \bibnamefont
  {White}},\ }\href {\doibase 10.1063/1.1508370} {\bibfield  {journal}
  {\bibinfo  {journal} {The Journal of Chemical Physics}\ }\textbf {\bibinfo
  {volume} {117}},\ \bibinfo {pages} {7472} (\bibinfo {year} {2002})},\ \Eprint
  {http://arxiv.org/abs/https://pubs.aip.org/aip/jcp/article-pdf/117/16/7472/19317208/7472\_1\_online.pdf}
  {https://pubs.aip.org/aip/jcp/article-pdf/117/16/7472/19317208/7472\_1\_online.pdf}
  \BibitemShut {NoStop}%
\bibitem [{\citenamefont {MacDonald}\ \emph {et~al.}(1988)\citenamefont
  {MacDonald}, \citenamefont {Girvin},\ and\ \citenamefont
  {Yoshioka}}]{MacD1988}%
  \BibitemOpen
  \bibfield  {author} {\bibinfo {author} {\bibfnamefont {A.~H.}\ \bibnamefont
  {MacDonald}}, \bibinfo {author} {\bibfnamefont {S.~M.}\ \bibnamefont
  {Girvin}}, \ and\ \bibinfo {author} {\bibfnamefont {D.}~\bibnamefont
  {Yoshioka}},\ }\href {\doibase 10.1103/PhysRevB.37.9753} {\bibfield
  {journal} {\bibinfo  {journal} {Physical Review B}\ }\textbf {\bibinfo
  {volume} {37}},\ \bibinfo {pages} {9753} (\bibinfo {year} {1988})},\ \bibinfo
  {note} {publisher: American Physical Society}\BibitemShut {NoStop}%
\bibitem [{\citenamefont {Anisimov}\ \emph {et~al.}(1991)\citenamefont
  {Anisimov}, \citenamefont {Zaanen},\ and\ \citenamefont
  {Andersen}}]{Anisimov:1991}%
  \BibitemOpen
  \bibfield  {author} {\bibinfo {author} {\bibfnamefont {V.~I.}\ \bibnamefont
  {Anisimov}}, \bibinfo {author} {\bibfnamefont {J.}~\bibnamefont {Zaanen}}, \
  and\ \bibinfo {author} {\bibfnamefont {O.~K.}\ \bibnamefont {Andersen}},\
  }\href {\doibase 10.1103/PhysRevB.44.943} {\bibfield  {journal} {\bibinfo
  {journal} {Phys. Rev. B}\ }\textbf {\bibinfo {volume} {44}},\ \bibinfo
  {pages} {943} (\bibinfo {year} {1991})}\BibitemShut {NoStop}%
\bibitem [{\citenamefont {Macke}\ \emph {et~al.}(2024)\citenamefont {Macke},
  \citenamefont {Timrov}, \citenamefont {Marzari},\ and\ \citenamefont
  {Ciacchi}}]{Marzari2024}%
  \BibitemOpen
  \bibfield  {author} {\bibinfo {author} {\bibfnamefont {E.}~\bibnamefont
  {Macke}}, \bibinfo {author} {\bibfnamefont {I.}~\bibnamefont {Timrov}},
  \bibinfo {author} {\bibfnamefont {N.}~\bibnamefont {Marzari}}, \ and\
  \bibinfo {author} {\bibfnamefont {L.~C.}\ \bibnamefont {Ciacchi}},\ }\href
  {\doibase 10.1021/acs.jctc.3c01403} {\bibfield  {journal} {\bibinfo
  {journal} {Journal of Chemical Theory and Computation}\ }\textbf {\bibinfo
  {volume} {20}},\ \bibinfo {pages} {4824–4843} (\bibinfo {year} {2024})},\
  \bibinfo {note} {pMID: 38820347},\ \Eprint
  {http://arxiv.org/abs/https://doi.org/10.1021/acs.jctc.3c01403}
  {https://doi.org/10.1021/acs.jctc.3c01403} \BibitemShut {NoStop}%
\bibitem [{\citenamefont {Georges}\ \emph {et~al.}(1996)\citenamefont
  {Georges}, \citenamefont {Kotliar}, \citenamefont {Krauth},\ and\
  \citenamefont {Rozenberg}}]{RevModPhys.68.13}%
  \BibitemOpen
  \bibfield  {author} {\bibinfo {author} {\bibfnamefont {A.}~\bibnamefont
  {Georges}}, \bibinfo {author} {\bibfnamefont {G.}~\bibnamefont {Kotliar}},
  \bibinfo {author} {\bibfnamefont {W.}~\bibnamefont {Krauth}}, \ and\ \bibinfo
  {author} {\bibfnamefont {M.~J.}\ \bibnamefont {Rozenberg}},\ }\href {\doibase
  10.1103/RevModPhys.68.13} {\bibfield  {journal} {\bibinfo  {journal} {Rev.
  Mod. Phys.}\ }\textbf {\bibinfo {volume} {68}},\ \bibinfo {pages} {13}
  (\bibinfo {year} {1996})}\BibitemShut {NoStop}%
\bibitem [{\citenamefont {Kotliar}\ \emph {et~al.}(2006)\citenamefont
  {Kotliar}, \citenamefont {Savrasov}, \citenamefont {Haule}, \citenamefont
  {Oudovenko}, \citenamefont {Parcollet},\ and\ \citenamefont
  {Marianetti}}]{Kotliar2006}%
  \BibitemOpen
  \bibfield  {author} {\bibinfo {author} {\bibfnamefont {G.}~\bibnamefont
  {Kotliar}}, \bibinfo {author} {\bibfnamefont {S.~Y.}\ \bibnamefont
  {Savrasov}}, \bibinfo {author} {\bibfnamefont {K.}~\bibnamefont {Haule}},
  \bibinfo {author} {\bibfnamefont {V.~S.}\ \bibnamefont {Oudovenko}}, \bibinfo
  {author} {\bibfnamefont {O.}~\bibnamefont {Parcollet}}, \ and\ \bibinfo
  {author} {\bibfnamefont {C.~A.}\ \bibnamefont {Marianetti}},\ }\href
  {\doibase 10.1103/revmodphys.78.865} {\bibfield  {journal} {\bibinfo
  {journal} {Rev. Mod. Phys.}\ }\textbf {\bibinfo {volume} {78}},\ \bibinfo
  {pages} {865} (\bibinfo {year} {2006})}\BibitemShut {NoStop}%
\bibitem [{\citenamefont {Vollhardt}(2019)}]{Vollhardt:2019}%
  \BibitemOpen
  \bibfield  {author} {\bibinfo {author} {\bibfnamefont {D.}~\bibnamefont
  {Vollhardt}},\ }\enquote {\bibinfo {title} {Dynamical mean-field theory of
  strongly correlated electron systems},}\ in\ \href {\doibase
  10.7566/JPSCP.30.011001} {\emph {\bibinfo {booktitle} {Proceedings of the
  International Conference on Strongly Correlated Electron Systems
  (SCES2019)}}}\ (\bibinfo {year} {2019})\ \Eprint
  {http://arxiv.org/abs/https://journals.jps.jp/doi/pdf/10.7566/JPSCP.30.011001}
  {https://journals.jps.jp/doi/pdf/10.7566/JPSCP.30.011001} \BibitemShut
  {NoStop}%
\bibitem [{\citenamefont {Aryasetiawan}\ \emph {et~al.}(2006)\citenamefont
  {Aryasetiawan}, \citenamefont {Karlsson}, \citenamefont {Jepsen},\ and\
  \citenamefont {Schonberger}}]{arya2006}%
  \BibitemOpen
  \bibfield  {author} {\bibinfo {author} {\bibfnamefont {F.}~\bibnamefont
  {Aryasetiawan}}, \bibinfo {author} {\bibfnamefont {K.}~\bibnamefont
  {Karlsson}}, \bibinfo {author} {\bibfnamefont {O.}~\bibnamefont {Jepsen}}, \
  and\ \bibinfo {author} {\bibfnamefont {U.}~\bibnamefont {Schonberger}},\
  }\href {\doibase 10.1103/PhysRevB.74.125106} {\bibfield  {journal} {\bibinfo
  {journal} {Physical Review B}\ }\textbf {\bibinfo {volume} {74}},\ \bibinfo
  {pages} {125106} (\bibinfo {year} {2006})},\ \bibinfo {note} {arXiv:
  cond-mat/0603138}\BibitemShut {NoStop}%
\bibitem [{\citenamefont {Springer}\ and\ \citenamefont
  {Aryasetiawan}(1998)}]{Springer:1998}%
  \BibitemOpen
  \bibfield  {author} {\bibinfo {author} {\bibfnamefont {M.}~\bibnamefont
  {Springer}}\ and\ \bibinfo {author} {\bibfnamefont {F.}~\bibnamefont
  {Aryasetiawan}},\ }\href {\doibase 10.1103/PhysRevB.57.4364} {\bibfield
  {journal} {\bibinfo  {journal} {Phys. Rev. B}\ }\textbf {\bibinfo {volume}
  {57}},\ \bibinfo {pages} {4364} (\bibinfo {year} {1998})}\BibitemShut
  {NoStop}%
\bibitem [{\citenamefont {Aryasetiawan}\ \emph {et~al.}(2004)\citenamefont
  {Aryasetiawan}, \citenamefont {Imada}, \citenamefont {Georges}, \citenamefont
  {Kotliar}, \citenamefont {Biermann},\ and\ \citenamefont
  {Lichtenstein}}]{RPA:2004}%
  \BibitemOpen
  \bibfield  {author} {\bibinfo {author} {\bibfnamefont {F.}~\bibnamefont
  {Aryasetiawan}}, \bibinfo {author} {\bibfnamefont {M.}~\bibnamefont {Imada}},
  \bibinfo {author} {\bibfnamefont {A.}~\bibnamefont {Georges}}, \bibinfo
  {author} {\bibfnamefont {G.}~\bibnamefont {Kotliar}}, \bibinfo {author}
  {\bibfnamefont {S.}~\bibnamefont {Biermann}}, \ and\ \bibinfo {author}
  {\bibfnamefont {A.~I.}\ \bibnamefont {Lichtenstein}},\ }\href {\doibase
  10.1103/PhysRevB.70.195104} {\bibfield  {journal} {\bibinfo  {journal} {Phys.
  Rev. B}\ }\textbf {\bibinfo {volume} {70}},\ \bibinfo {pages} {195104}
  (\bibinfo {year} {2004})}\BibitemShut {NoStop}%
\bibitem [{\citenamefont {Karlsson}\ \emph {et~al.}(2010)\citenamefont
  {Karlsson}, \citenamefont {Aryasetiawan},\ and\ \citenamefont
  {Jepsen}}]{Jepsen:2010}%
  \BibitemOpen
  \bibfield  {author} {\bibinfo {author} {\bibfnamefont {K.}~\bibnamefont
  {Karlsson}}, \bibinfo {author} {\bibfnamefont {F.}~\bibnamefont
  {Aryasetiawan}}, \ and\ \bibinfo {author} {\bibfnamefont {O.}~\bibnamefont
  {Jepsen}},\ }\href {\doibase 10.1103/PhysRevB.81.245113} {\bibfield
  {journal} {\bibinfo  {journal} {Phys. Rev. B}\ }\textbf {\bibinfo {volume}
  {81}},\ \bibinfo {pages} {245113} (\bibinfo {year} {2010})}\BibitemShut
  {NoStop}%
\bibitem [{\citenamefont {Honerkamp}\ \emph {et~al.}(2018)\citenamefont
  {Honerkamp}, \citenamefont {Shinaoka}, \citenamefont {Assaad},\ and\
  \citenamefont {Werner}}]{Hohnerkamp:2018}%
  \BibitemOpen
  \bibfield  {author} {\bibinfo {author} {\bibfnamefont {C.}~\bibnamefont
  {Honerkamp}}, \bibinfo {author} {\bibfnamefont {H.}~\bibnamefont {Shinaoka}},
  \bibinfo {author} {\bibfnamefont {F.~F.}\ \bibnamefont {Assaad}}, \ and\
  \bibinfo {author} {\bibfnamefont {P.}~\bibnamefont {Werner}},\ }\href
  {\doibase 10.1103/PhysRevB.98.235151} {\bibfield  {journal} {\bibinfo
  {journal} {Phys. Rev. B}\ }\textbf {\bibinfo {volume} {98}},\ \bibinfo
  {pages} {235151} (\bibinfo {year} {2018})}\BibitemShut {NoStop}%
\bibitem [{\citenamefont {Wannier}(1937)}]{Wannier:1937}%
  \BibitemOpen
  \bibfield  {author} {\bibinfo {author} {\bibfnamefont {G.~H.}\ \bibnamefont
  {Wannier}},\ }\href {\doibase 10.1103/PhysRev.52.191} {\bibfield  {journal}
  {\bibinfo  {journal} {Phys. Rev.}\ }\textbf {\bibinfo {volume} {52}},\
  \bibinfo {pages} {191} (\bibinfo {year} {1937})}\BibitemShut {NoStop}%
\bibitem [{\citenamefont {Kohn}(1959)}]{Kohn:1959}%
  \BibitemOpen
  \bibfield  {author} {\bibinfo {author} {\bibfnamefont {W.}~\bibnamefont
  {Kohn}},\ }\href {\doibase 10.1103/PhysRev.115.809} {\bibfield  {journal}
  {\bibinfo  {journal} {Phys. Rev.}\ }\textbf {\bibinfo {volume} {115}},\
  \bibinfo {pages} {809} (\bibinfo {year} {1959})}\BibitemShut {NoStop}%
\bibitem [{\citenamefont {Marzari}\ \emph {et~al.}(2012)\citenamefont
  {Marzari}, \citenamefont {Mostofi}, \citenamefont {Yates}, \citenamefont
  {Souza},\ and\ \citenamefont {Vanderbilt}}]{RevModPhys.84.1419}%
  \BibitemOpen
  \bibfield  {author} {\bibinfo {author} {\bibfnamefont {N.}~\bibnamefont
  {Marzari}}, \bibinfo {author} {\bibfnamefont {A.~A.}\ \bibnamefont
  {Mostofi}}, \bibinfo {author} {\bibfnamefont {J.~R.}\ \bibnamefont {Yates}},
  \bibinfo {author} {\bibfnamefont {I.}~\bibnamefont {Souza}}, \ and\ \bibinfo
  {author} {\bibfnamefont {D.}~\bibnamefont {Vanderbilt}},\ }\href {\doibase
  10.1103/RevModPhys.84.1419} {\bibfield  {journal} {\bibinfo  {journal} {Rev.
  Mod. Phys.}\ }\textbf {\bibinfo {volume} {84}},\ \bibinfo {pages} {1419}
  (\bibinfo {year} {2012})}\BibitemShut {NoStop}%
\bibitem [{\citenamefont {Liechtenstein}\ \emph {et~al.}(1987)\citenamefont
  {Liechtenstein}, \citenamefont {Katsnelson}, \citenamefont {Antropov},\ and\
  \citenamefont {Gubanov}}]{Liechtenstein:1987}%
  \BibitemOpen
  \bibfield  {author} {\bibinfo {author} {\bibfnamefont {A.}~\bibnamefont
  {Liechtenstein}}, \bibinfo {author} {\bibfnamefont {M.}~\bibnamefont
  {Katsnelson}}, \bibinfo {author} {\bibfnamefont {V.}~\bibnamefont
  {Antropov}}, \ and\ \bibinfo {author} {\bibfnamefont {V.}~\bibnamefont
  {Gubanov}},\ }\href {\doibase 10.1016/0304-8853(87)90721-9} {\bibfield
  {journal} {\bibinfo  {journal} {Journal of Magnetism and Magnetic Materials}\
  }\textbf {\bibinfo {volume} {67}},\ \bibinfo {pages} {65–74} (\bibinfo
  {year} {1987})}\BibitemShut {NoStop}%
\bibitem [{\citenamefont {He}\ \emph {et~al.}(2021)\citenamefont {He},
  \citenamefont {Helbig}, \citenamefont {Verstraete},\ and\ \citenamefont
  {Bousquet}}]{TB2J:2021}%
  \BibitemOpen
  \bibfield  {author} {\bibinfo {author} {\bibfnamefont {X.}~\bibnamefont
  {He}}, \bibinfo {author} {\bibfnamefont {N.}~\bibnamefont {Helbig}}, \bibinfo
  {author} {\bibfnamefont {M.~J.}\ \bibnamefont {Verstraete}}, \ and\ \bibinfo
  {author} {\bibfnamefont {E.}~\bibnamefont {Bousquet}},\ }\href {\doibase
  10.1016/j.cpc.2021.107938} {\bibfield  {journal} {\bibinfo  {journal}
  {Computer Physics Communications}\ }\textbf {\bibinfo {volume} {264}},\
  \bibinfo {pages} {107938} (\bibinfo {year} {2021})}\BibitemShut {NoStop}%
\bibitem [{\citenamefont {Penrose}(1956)}]{Penr1956}%
  \BibitemOpen
  \bibfield  {author} {\bibinfo {author} {\bibfnamefont {R.}~\bibnamefont
  {Penrose}},\ }\href {\doibase 10.1017/S0305004100030929} {\bibfield
  {journal} {\bibinfo  {journal} {Mathematical Proceedings of the Cambridge
  Philosophical Society}\ }\textbf {\bibinfo {volume} {52}},\ \bibinfo {pages}
  {17} (\bibinfo {year} {1956})}\BibitemShut {NoStop}%
\bibitem [{\citenamefont {Helgaker}\ \emph {et~al.}(2000)\citenamefont
  {Helgaker}, \citenamefont {Jorgensen},\ and\ \citenamefont
  {Olsen}}]{Helg2000}%
  \BibitemOpen
  \bibfield  {author} {\bibinfo {author} {\bibfnamefont {T.}~\bibnamefont
  {Helgaker}}, \bibinfo {author} {\bibfnamefont {P.}~\bibnamefont {Jorgensen}},
  \ and\ \bibinfo {author} {\bibfnamefont {J.}~\bibnamefont {Olsen}},\ }\href
  {\doibase 10.1002/9781119019572.ch11} {\emph {\bibinfo {title} {Molecular
  {Electronic}-{Structure} {Theory}}}}\ (\bibinfo  {publisher} {John Wiley \&
  Sons, Ltd},\ \bibinfo {year} {2000})\ pp.\ \bibinfo {pages}
  {523--597}\BibitemShut {NoStop}%
\bibitem [{\citenamefont {Löwdin}(1955)}]{loew1955}%
  \BibitemOpen
  \bibfield  {author} {\bibinfo {author} {\bibfnamefont {P.-O.}\ \bibnamefont
  {Löwdin}},\ }\href {\doibase 10.1103/PhysRev.97.1474} {\bibfield  {journal}
  {\bibinfo  {journal} {Physical Review}\ }\textbf {\bibinfo {volume} {97}},\
  \bibinfo {pages} {1474} (\bibinfo {year} {1955})},\ \bibinfo {note}
  {publisher: American Physical Society}\BibitemShut {NoStop}%
\bibitem [{\citenamefont {Lindh}\ and\ \citenamefont
  {Persson}(1994)}]{Lind1994}%
  \BibitemOpen
  \bibfield  {author} {\bibinfo {author} {\bibfnamefont {R.}~\bibnamefont
  {Lindh}}\ and\ \bibinfo {author} {\bibfnamefont {B.~J.}\ \bibnamefont
  {Persson}},\ }\href {\doibase 10.1021/ja00090a047} {\bibfield  {journal}
  {\bibinfo  {journal} {Journal of the American Chemical Society}\ }\textbf
  {\bibinfo {volume} {116}},\ \bibinfo {pages} {4963} (\bibinfo {year}
  {1994})}\BibitemShut {NoStop}%
\bibitem [{\citenamefont {Salem}\ and\ \citenamefont
  {Rowland}(1972)}]{Sale1972}%
  \BibitemOpen
  \bibfield  {author} {\bibinfo {author} {\bibfnamefont {L.}~\bibnamefont
  {Salem}}\ and\ \bibinfo {author} {\bibfnamefont {C.}~\bibnamefont
  {Rowland}},\ }\href {\doibase https://doi.org/10.1002/anie.197200921}
  {\bibfield  {journal} {\bibinfo  {journal} {Angewandte Chemie International
  Edition in English}\ }\textbf {\bibinfo {volume} {11}},\ \bibinfo {pages}
  {92} (\bibinfo {year} {1972})}\BibitemShut {NoStop}%
\bibitem [{\citenamefont {McWeeny}\ and\ \citenamefont
  {Coulson}(1997)}]{McWe1957}%
  \BibitemOpen
  \bibfield  {author} {\bibinfo {author} {\bibfnamefont {R.}~\bibnamefont
  {McWeeny}}\ and\ \bibinfo {author} {\bibfnamefont {C.~A.}\ \bibnamefont
  {Coulson}},\ }\href {\doibase 10.1098/rspa.1957.0125} {\bibfield  {journal}
  {\bibinfo  {journal} {Proceedings of the Royal Society of London. Series A.
  Mathematical and Physical Sciences}\ }\textbf {\bibinfo {volume} {241}},\
  \bibinfo {pages} {239} (\bibinfo {year} {1997})},\ \bibinfo {note}
  {publisher: Royal Society}\BibitemShut {NoStop}%
\bibitem [{\citenamefont {Roos}\ \emph {et~al.}(1980)\citenamefont {Roos},
  \citenamefont {Taylor},\ and\ \citenamefont {Sigbahn}}]{Roos1980}%
  \BibitemOpen
  \bibfield  {author} {\bibinfo {author} {\bibfnamefont {B.~O.}\ \bibnamefont
  {Roos}}, \bibinfo {author} {\bibfnamefont {P.~R.}\ \bibnamefont {Taylor}}, \
  and\ \bibinfo {author} {\bibfnamefont {P.~E.}\ \bibnamefont {Sigbahn}},\
  }\href {\doibase 10.1016/0301-0104(80)80045-0} {\bibfield  {journal}
  {\bibinfo  {journal} {Chemical Physics}\ }\textbf {\bibinfo {volume} {48}},\
  \bibinfo {pages} {157–173} (\bibinfo {year} {1980})}\BibitemShut {NoStop}%
\bibitem [{\citenamefont {Lawley}(2009)}]{Lawl2009}%
  \BibitemOpen
  \bibfield  {author} {\bibinfo {author} {\bibfnamefont {K.~P.}\ \bibnamefont
  {Lawley}},\ }\href@noop {} {\emph {\bibinfo {title} {Ab Initio Methods in
  Quantum Chemistry, Volume 69, Part 2}}}\ (\bibinfo  {publisher} {John Wiley
  \& Sons},\ \bibinfo {year} {2009})\BibitemShut {NoStop}%
\bibitem [{\citenamefont {Anderson}(1961)}]{Ande1961}%
  \BibitemOpen
  \bibfield  {author} {\bibinfo {author} {\bibfnamefont {P.~W.}\ \bibnamefont
  {Anderson}},\ }\href {\doibase 10.1103/PhysRev.124.41} {\bibfield  {journal}
  {\bibinfo  {journal} {Physical Review}\ }\textbf {\bibinfo {volume} {124}},\
  \bibinfo {pages} {41} (\bibinfo {year} {1961})},\ \bibinfo {note} {publisher:
  American Physical Society}\BibitemShut {NoStop}%
\bibitem [{\citenamefont {Kehrein}\ and\ \citenamefont
  {Mielke}(1996)}]{Kehr1996}%
  \BibitemOpen
  \bibfield  {author} {\bibinfo {author} {\bibfnamefont {S.~K.}\ \bibnamefont
  {Kehrein}}\ and\ \bibinfo {author} {\bibfnamefont {A.}~\bibnamefont
  {Mielke}},\ }\href {\doibase 10.1006/aphy.1996.0123} {\bibfield  {journal}
  {\bibinfo  {journal} {Annals of Physics}\ }\textbf {\bibinfo {volume}
  {252}},\ \bibinfo {pages} {1} (\bibinfo {year} {1996})}\BibitemShut {NoStop}%
\bibitem [{\citenamefont {Haq}\ and\ \citenamefont {Singh}(2020)}]{HaqS2020}%
  \BibitemOpen
  \bibfield  {author} {\bibinfo {author} {\bibfnamefont {R.~U.}\ \bibnamefont
  {Haq}}\ and\ \bibinfo {author} {\bibfnamefont {K.}~\bibnamefont {Singh}},\
  }\href {http://arxiv.org/abs/2004.06534} {\bibfield  {journal} {\bibinfo
  {journal} {arXiv:2004.06534 [cond-mat]}\ } (\bibinfo {year} {2020})},\
  \bibinfo {note} {arXiv: 2004.06534}\BibitemShut {NoStop}%
\bibitem [{\citenamefont {Hewson}(1997)}]{Hews1997}%
  \BibitemOpen
  \bibfield  {author} {\bibinfo {author} {\bibfnamefont {A.~C.}\ \bibnamefont
  {Hewson}},\ }\href@noop {} {\emph {\bibinfo {title} {The Kondo Problem to
  Heavy Fermions}}}\ (\bibinfo  {publisher} {Cambridge University Press},\
  \bibinfo {year} {1997})\BibitemShut {NoStop}%
\bibitem [{\citenamefont {Hubbard}(1963)}]{Hubb1963}%
  \BibitemOpen
  \bibfield  {author} {\bibinfo {author} {\bibfnamefont {J.}~\bibnamefont
  {Hubbard}},\ }\href@noop {} {\bibfield  {journal} {\bibinfo  {journal}
  {{Proceedings} of the {Royal} {Society} of {London}. {Series} {A}.
  {Mathematical} and {Physical} {Sciences}}\ ,\ \bibinfo {pages} {238}}
  (\bibinfo {year} {1963})}\BibitemShut {NoStop}%
\bibitem [{\citenamefont {Essler}\ \emph {et~al.}(2005)\citenamefont {Essler},
  \citenamefont {Frahm}, \citenamefont {Göhmann}, \citenamefont {Klümper},\
  and\ \citenamefont {Korepin}}]{Essl2005}%
  \BibitemOpen
  \bibfield  {author} {\bibinfo {author} {\bibfnamefont {F.~H.~L.}\
  \bibnamefont {Essler}}, \bibinfo {author} {\bibfnamefont {H.}~\bibnamefont
  {Frahm}}, \bibinfo {author} {\bibfnamefont {F.}~\bibnamefont {Göhmann}},
  \bibinfo {author} {\bibfnamefont {A.}~\bibnamefont {Klümper}}, \ and\
  \bibinfo {author} {\bibfnamefont {V.~E.}\ \bibnamefont {Korepin}},\
  }\href@noop {} {\emph {\bibinfo {title} {The {One}-{Dimensional} {Hubbard}
  {Model}}}}\ (\bibinfo  {publisher} {Cambridge University Press},\ \bibinfo
  {year} {2005})\BibitemShut {NoStop}%
\bibitem [{\citenamefont {Gong}\ \emph {et~al.}(2017)\citenamefont {Gong},
  \citenamefont {Li}, \citenamefont {Li}, \citenamefont {Ji}, \citenamefont
  {Stern}, \citenamefont {Xia}, \citenamefont {Cao}, \citenamefont {Bao},
  \citenamefont {Wang}, \citenamefont {Wang}, \citenamefont {Qiu},
  \citenamefont {Cava}, \citenamefont {Louie}, \citenamefont {Xia},\ and\
  \citenamefont {Zhang}}]{Gong2017}%
  \BibitemOpen
  \bibfield  {author} {\bibinfo {author} {\bibfnamefont {C.}~\bibnamefont
  {Gong}}, \bibinfo {author} {\bibfnamefont {L.}~\bibnamefont {Li}}, \bibinfo
  {author} {\bibfnamefont {Z.}~\bibnamefont {Li}}, \bibinfo {author}
  {\bibfnamefont {H.}~\bibnamefont {Ji}}, \bibinfo {author} {\bibfnamefont
  {A.}~\bibnamefont {Stern}}, \bibinfo {author} {\bibfnamefont
  {Y.}~\bibnamefont {Xia}}, \bibinfo {author} {\bibfnamefont {T.}~\bibnamefont
  {Cao}}, \bibinfo {author} {\bibfnamefont {W.}~\bibnamefont {Bao}}, \bibinfo
  {author} {\bibfnamefont {C.}~\bibnamefont {Wang}}, \bibinfo {author}
  {\bibfnamefont {Y.}~\bibnamefont {Wang}}, \bibinfo {author} {\bibfnamefont
  {Z.~Q.}\ \bibnamefont {Qiu}}, \bibinfo {author} {\bibfnamefont {R.~J.}\
  \bibnamefont {Cava}}, \bibinfo {author} {\bibfnamefont {S.~G.}\ \bibnamefont
  {Louie}}, \bibinfo {author} {\bibfnamefont {J.}~\bibnamefont {Xia}}, \ and\
  \bibinfo {author} {\bibfnamefont {X.}~\bibnamefont {Zhang}},\ }\href
  {\doibase 10.1038/nature22060} {\bibfield  {journal} {\bibinfo  {journal}
  {Nature}\ }\textbf {\bibinfo {volume} {546}},\ \bibinfo {pages} {265}
  (\bibinfo {year} {2017})}\BibitemShut {NoStop}%
\bibitem [{\citenamefont {Huang}\ \emph {et~al.}(2017)\citenamefont {Huang},
  \citenamefont {Clark}, \citenamefont {Navarro-Moratalla}, \citenamefont
  {Klein}, \citenamefont {Cheng}, \citenamefont {Seyler}, \citenamefont
  {Zhong}, \citenamefont {Schmidgall}, \citenamefont {McGuire}, \citenamefont
  {Cobden}, \citenamefont {Yao}, \citenamefont {Xiao}, \citenamefont
  {Jarillo-Herrero},\ and\ \citenamefont {Xu}}]{Huang2017}%
  \BibitemOpen
  \bibfield  {author} {\bibinfo {author} {\bibfnamefont {B.}~\bibnamefont
  {Huang}}, \bibinfo {author} {\bibfnamefont {G.}~\bibnamefont {Clark}},
  \bibinfo {author} {\bibfnamefont {E.}~\bibnamefont {Navarro-Moratalla}},
  \bibinfo {author} {\bibfnamefont {D.~R.}\ \bibnamefont {Klein}}, \bibinfo
  {author} {\bibfnamefont {R.}~\bibnamefont {Cheng}}, \bibinfo {author}
  {\bibfnamefont {K.~L.}\ \bibnamefont {Seyler}}, \bibinfo {author}
  {\bibfnamefont {D.}~\bibnamefont {Zhong}}, \bibinfo {author} {\bibfnamefont
  {E.}~\bibnamefont {Schmidgall}}, \bibinfo {author} {\bibfnamefont {M.~A.}\
  \bibnamefont {McGuire}}, \bibinfo {author} {\bibfnamefont {D.~H.}\
  \bibnamefont {Cobden}}, \bibinfo {author} {\bibfnamefont {W.}~\bibnamefont
  {Yao}}, \bibinfo {author} {\bibfnamefont {D.}~\bibnamefont {Xiao}}, \bibinfo
  {author} {\bibfnamefont {P.}~\bibnamefont {Jarillo-Herrero}}, \ and\ \bibinfo
  {author} {\bibfnamefont {X.}~\bibnamefont {Xu}},\ }\href {\doibase
  10.1038/nature22391} {\bibfield  {journal} {\bibinfo  {journal} {Nature}\
  }\textbf {\bibinfo {volume} {546}},\ \bibinfo {pages} {270} (\bibinfo {year}
  {2017})}\BibitemShut {NoStop}%
\bibitem [{\citenamefont {Esteras}\ and\ \citenamefont
  {Baldoví}(2023)}]{Esteras2023}%
  \BibitemOpen
  \bibfield  {author} {\bibinfo {author} {\bibfnamefont {D.~L.}\ \bibnamefont
  {Esteras}}\ and\ \bibinfo {author} {\bibfnamefont {J.~J.}\ \bibnamefont
  {Baldoví}},\ }\href {\doibase https://doi.org/10.1016/j.mtelec.2023.100072}
  {\bibfield  {journal} {\bibinfo  {journal} {Materials Today Electronics}\
  }\textbf {\bibinfo {volume} {6}},\ \bibinfo {pages} {100072} (\bibinfo {year}
  {2023})}\BibitemShut {NoStop}%
\bibitem [{\citenamefont {{HQS Quantumsimulations GmbH}}(2024)}]{asf2024}%
  \BibitemOpen
  \bibfield  {author} {\bibinfo {author} {\bibnamefont {{HQS Quantumsimulations
  GmbH}}},\ }\href {https://github.com/HQSquantumsimulations/ActiveSpaceFinder}
  {\enquote {\bibinfo {title} {{ActiveSpaceFinder}},}\ } (\bibinfo {year}
  {2024}),\ \Eprint
  {http://arxiv.org/abs/github.com/HQSquantumsimulations/ActiveSpaceFinder}
  {github.com/HQSquantumsimulations/ActiveSpaceFinder} \BibitemShut {NoStop}%
\end{thebibliography}%
\end{document}